\documentclass[pdflatex,sn-mathphys-num]{sn-jnl}

\makeatletter
\providecommand{\@outputbox@removebskip}{} 
\let\MFL@makecol\@makecol                  
\makeatother

\usepackage{graphicx}%
\usepackage{multirow}%
\usepackage{amsmath,amssymb,amsfonts}%
\usepackage{amsthm}%
\usepackage{mathrsfs}%
\usepackage[title]{appendix}%
\usepackage{xcolor}%
\usepackage{textcomp}%
\usepackage{manyfoot}%
\usepackage{booktabs}%
\usepackage{algorithm}%
\usepackage{algorithmicx}%
\usepackage{algpseudocode}%
\usepackage{listings}%
\usepackage{mathrsfs}
\usepackage{lineno}  
\usepackage{multicol}

\theoremstyle{thmstyleone}%
\newtheorem{theorem}{Theorem}
\newtheorem{proposition}[theorem]{Proposition}%

\theoremstyle{thmstyletwo}%
\newtheorem{example}{Example}%
\newtheorem{remark}{Remark}%

\theoremstyle{thmstylethree}%

\begin{document}
	
	\title[Article Title]{A Relativity-Based Framework for Statistical Testing Guided by the Independence of Ancillary Statistics: Methodology and Nonparametric Illustrations}

	
	\author*[1]{\fnm{Albert} \sur{Vexler}}\email{avexler@buffalo.edu}
	
	\author[1]{\fnm{Douglas} \sur{ Landsittel}}\email{dplansit@buffalo.edu}

	\affil*[1]{\orgdiv{Department of Biostatistics}, \orgname{The State University of New York at Buffalo}, \orgaddress{ \city{Buffalo}, \postcode{14214}, \state{NY}, \country{USA}}}
	
	
	\abstract{
		A test statistic is typically constructed to discriminate effectively between competing hypotheses.
		In contrast, we propose and examine a framework that shifts attention to ancillary statistics—quantities whose distributions remain invariant under the tested hypotheses.	
		Rather than directly optimizing discriminatory power, the proposed approach seeks to construct test statistics that
		exhibit relative independence from ancillary structures.
		We show that reducing the dependence between a test statistic and a vector of ancillary statistics can yield the most powerful (MP) decision-making procedure.
		We establish a Basu-type independence result and show that certain forms of MP test statistics characterize the underlying distribution.	
		These principles are developed through decision-theoretic arguments and illustrated in two nonparametric applications.
		Ancillary-guided modifications of the Shapiro–Wilk, Anderson–Darling, Cram\'{e}r-von Mises, and Kolmogorov–Smirnov tests 
		deliver twofold efficiency gains under symmetric alternatives. In multivariate mean testing, a simple trace-normalized statistic reduces ancillary dependence and then outperforms Hotelling’s procedure under heavy-tailed distributions, while the classical test remains optimal under normality. 
		The proposed framework is simple to implement and provides a theoretically grounded strategy for enhancing the power of statistical tests in practice.}

	
	
	
	\maketitle

\section{Introduction}\label{Intro}
The development of decision-making mechanisms lies at the heart of statistical theory and underpins a wide range of inferential procedures in practice. 
This paper explores a  strategy for constructing, improving,  and comparing test statistics, emphasizing performance gains achieved through structural modifications guided by the principle of independence from ancillary statistics.

To illustrate the ideas of this paper, consider a simple example. Suppose we observe a sample of \( n \) independent and identically distributed (i.i.d.) data points \( X_1, \ldots, X_n \) with variance $1$, and our goal is to test the hypothesis \( H_0 : \mathrm{E}(X_1) = 0 \) versus \( H_1 : \mathrm{E}(X_1) > 0 \). A natural choice of test statistic is based on the sample mean \( \bar{X}_n = \sum_{i=1}^n X_i / n \), which discriminates between the hypotheses via the rule: we reject \( H_0 \) if \( n^{0.5} \bar{X}_n > C \), where \( C \) is a test threshold.
The law of large numbers and the central limit theorem justify this choice by ensuring that \( \bar{X}_n \) converges to its expectation and has asymptotically distinct distributions under \( H_0 \) and \( H_1 \). This reflects a classical strategy: selecting a statistic with strong discriminatory power.

Alternatively, within a relativity-based framework, one may focus on constructing or modifying test statistics to be (exactly or approximately) independent of certain ancillary quantities---statistics whose distributions are invariant (or approximately invariant) under the null and alternative. For instance, continuing with the example above,  the sample variance, 
$S^2_n=\sum_{i=1}^n(X_i-\bar{X}_n)^2/n,$ is ancillary in this context. By seeking a test statistic that is independent of \( S_n^2 \), one can develop or refine procedures that exhibit improved performance.
More broadly, this idea can guide both the construction of new test statistics and the modification of existing ones, by reducing their dependence on a relevant set of ancillary statistics.
For example, modifying the sample mean to reduce its dependence on the sample variance can yield a more powerful test when the distribution of \( X_1 \) is skewed \citep{vexler2024characterization}. It is worth noting that, when \( X_1 \) follows a normal distribution, the statistic \( \bar{X}_n \) is independent of the sample variance \( S^2_n \), and can represent a most powerful  test statistic in the Neyman--Pearson sense.

We use the term "relativity" in the following sense. A standard route to powerful decisions is to seek statistics with relevant maximal discriminating ability. Our results show that, when the conditions identified in this paper are satisfied, an apparently opposite approach—constructing a system of statistics that is “powerless”—can become relatively the same as a maximal-discrimination–based rule. Informally, creating a best-discriminating statistic and developing a system of powerless statistics may be viewed as two equivalent representations of the same inferential content. The connection to relativity theory is analogical: the ancillary system can be viewed as a reference frame, and once inference is expressed relative to this frame, the resulting conclusion is stable across different but equivalent constructions.

That is, by identifying a collection of statistics with no discriminatory ability---denote this set by \( A \)---one can construct a function of the 
data that is independent 
of \( A \), and thereby obtain a statistic that is highly efficient for testing the hypotheses of interest.

In this context, an intriguing question arises: how should the set \( A \) be chosen to ensure maximal testing efficiency? This leads to several related inquiries. For example, can we define \( A \) so that a statistic independent of \( A \) yields a most powerful test? In practice, if several ancillary statistics \( A_1, \ldots, A_k \) are available, which among them should be used to guide the transformation of a test statistic? Should \( A \) be selected to characterize the data distributions under the null and alternative hypotheses, or chosen as a maximal ancillary statistic in some formal sense? These questions are examined in this research, but many relevant aspects remain open for further investigation in future research.


Recently, \cite{vexler2024characterization} showed that transforming a test statistic \( T \)
to achieve independence from a relevant ancillary statistic \( A \) can yield a new statistic \( T_N \)
with greater power than the original \( T \). In this paper, we associate the term "Most Powerful (MP) Test" with ancillarity by defining a class of relevant ancillary statistics such that independence between a test statistic and this class plays a central role in achieving efficient test performance.

We adopt a theoretical formalism in which, under the considered decision-making structure, an MP mechanism exists—for example, the Neyman-Pearson likelihood ratio test. In many practical settings, however, an  MP test may be unavailable. In such cases, we interpret the term "MP" as a theoretical ideal: a target that can be  closely approached by approximating the underlying assumptions and structural principles discussed in the methodological sections of this paper. As an application, we consider nonparametric tests for normality that exemplify this perspective.

\begin{remark} [Relation to conditioning on ancillaries, invariance, and robustness.]
\emph{ A common theme in the ancillary-statistics literature is to use ancillarity primarily for
\emph{calibration}: one conditions on (exact or approximate) ancillary information to
remove nuisance-driven variability and obtain sharper conditional likelihood/pivot
approximations for parametric inference, yielding a conditional analysis rather than a
general recipe for modifying an existing test statistic \citep{Fraser2004_Ancillaries, Gh2}.
\\
A closely related reduction principle in hypothesis testing is \emph{invariance}: when a
transformation group leaves the testing problem unchanged, one restricts attention to
group-invariant decision rules (equivalently, functions of a maximal invariant), which
leads to exact randomization-based tests under the induced transformation distribution;
recent work shows that power can depend materially on how the acting (sub)group is
chosen \citep{KoningHemerik2024_RepSubgroup}.
\\
Robust testing pursues a different objective---stability of level and power under outliers,
heavy tails, or mild misspecification---typically by robustifying score/Wald/likelihood
constructions or by using divergence-based criteria \citep{HeritierRonchetti1994_BoundedInfluenceTests, CantoniRonchetti2001_RobustGLM, TomaBroniatowski2011_DualDivergence}.
\\
The present paper takes a complementary, \emph{test-construction} perspective: we treat
ancillary structure not mainly as something to condition on (calibration) or to quotient
out by symmetry (invariance), but as a design constraint - build or adjust a test so that
the decision rule is as insensitive as possible to ancillary variation.
This viewpoint is extended to include a characterization of the MP ideal (MP test
procedures), providing a conceptual benchmark based on independence from ancillary
structures and identifying conditions under which such a characterization holds.
This perspective leads to concrete construction and modification rules that reduce
``ancillary leakage'' - so that ancillary variation has minimal impact on the rejection
decision - thereby concentrating power on features that distinguish the competing
hypotheses in the regimes targeted by the paper. This, for example, creates an
opportunity to present nonparametric applications of the proposed technique.}
\end{remark}

Ancillarity underpins many foundational  statistical concepts \citep{lehmann1992ancillarity, Gh2}. Perhaps one of the most widely recognized results in the study of ancillary statistics is Basu's theorem \citep{basu}.  In simple terms, Basu's theorem states that if one has a complete sufficient statistic---containing all the information about an unknown parameter---and an ancillary statistic---containing no information about the parameter---then these two statistics are independent.


It appears that we are particularly interested 
in a type of conversion of Basu's result. However, it is important to note that a direct or naïve reversal of the classical Basu theorem---e.g., concluding sufficiency from independence---is generally false \citep{kagan2013nile}.

This paper deals with an extension of Basu's theorem and its conversion-type results to decision-making fields. We propose and examine the relativity-based framework, showing relevant examples. A consequence of this perspective is characterizations of the normal and negative exponential distributions. In analogy to these results, we refer the reader to~\cite{pfanzagl1968characterization} for an example of characterizing the one-parameter exponential family through the existence of MP test statistics.

We emphasize the practical value of the proposed methodology in goodness-of-fit and multivariate mean testing. In nonparametric settings, the main results developed in this study may be applied approximately, in much the same spirit as the Neyman–Pearson framework offers guidance for constructing test statistics beyond strict parametric assumptions. In both applications, the method is simple, transparent, and efficient.

\begin{remark}[Approximate MP benchmark and ancillary independence.]\label{Remark2:aprAnc}
\emph{ 
Throughout this paper we distinguish between exact independence (as in the formal results of Sections~\ref{s1} and~\ref{sc2}) and approximate independence, which is what is typically achievable in nonparametric settings or when nuisance components must be estimated. We also distinguish exact ancillarity from approximate ancillarity (e.g., \citet{skovgaard1985second}), where a statistic has approximately the same distribution under the null and alternative hypotheses. By “approximate independence” we mean that the dependence between the proposed test statistic and the (exactly or approximately) ancillary component is small (as can be assessed by a dependence measure) or vanishes in an appropriate asymptotic regime under both hypotheses. In such cases, the MP property is used only as a conceptual benchmark: reducing ancillary dependence is a design constraint intended to mitigate ancillary leakage and improve power, but it does not by itself guarantee exact 
optimality.
Accordingly, when only approximate ancillarity or approximate independence is available, the MP conclusions of Sections~\ref{s1} and~\ref{sc2}  are used only as  a guiding ideal leading to power improvements, not as a claim of exact optimality.
}
\end{remark}

The remainder of the paper is organized as follows. Section~\ref{s1} introduces the necessary notation and preliminaries. Section~\ref{sc2} develops a decision-theoretic approach that motivates the construction of test statistics with reduced dependence on ancillary structures and establishes conditions under which such procedures are MP. Moreover, a result in the spirit of Basu's theorem is reformulated and proved within a decision-theoretic context. The theoretical implications are illustrated through Examples~\ref{ex1}-\ref{ex3}.
We also demonstrate that certain forms of MP test statistics implicitly characterize the underlying data distribution.
Section~\ref{sc3} presents nonparametric applications, including modifications of classical tests for normality guided by ancillary-independence principles and a multivariate mean testing example in which Hotelling’s procedure is compared with a simple trace-normalized test.
Section~\ref{conclud} concludes with a discussion of methodological implications and potential directions for future research. Proofs of the main results are given in the Appendix, and additional technical derivations, along with several auxiliary remarks, are provided in the online Supplement.
\section{Some notation and preliminaries}\label{s1} 
Without loss of generality and to simplify the exposition of the main objectives of this paper, we introduce the following formal notation.

Assume we observe data denoted by \(D\), and our goal is to test a null hypothesis \(H_0\) against an alternative \(H_1\). Let \(T = T(D)\) be a real-valued test statistic constructed from \(D\), such that the hypothesis \(H_0\) is rejected when \(T(D) > C\), for some fixed threshold \(C\). We denote the probability and expectation under \(H_k\) by \(\Pr_k\) and \(\mathrm{E}_k\), respectively, for \(k \in \{0, 1\}\).

Throughout this paper, we assume that the distributions \(\Pr_0\) and \(\Pr_1\) of \(D\) are absolutely continuous with respect to a \(\sigma\)-finite measure \(\tau\), defined over a class \(\Upsilon\) of measurable subsets of a space \(\mathbb{X}\), in which \(D\) takes values. Then, there exist nonnegative generalized density functions \(f_0\) and \(f_1\) with respect to \(\tau\) such that for all \(\upsilon \in \Upsilon\) and \(k \in \{0, 1\}\),
$
{\Pr}_k(D \in \upsilon) = \int_{\upsilon} f_k(x) \, d\tau(x).
$
Note that \(f_0\) and \(f_1\) need not belong to the same parametric family of distributions.

Define \(f^T_0\) and \(f^T_1\) as the probability density functions of the statistic \(T(D)\) under \(H_0\) and \(H_1\), respectively. Then, for each \(k \in \{0, 1\}\) and any threshold \(t \in \mathbb{R}^1\),
\[
{\Pr}_k\{T(D) \le t\} = \int_{\mathbb{X}} I\{T(x) \le t\} f_k(x) \, d\tau(x) = \int_{-\infty}^t f^T_k(u) \, du,
\]
where \(I(\cdot)\) denotes the indicator function.  

In this setting, we assume that there exists some most powerful (MP) test statistic, denoted by \( \Lambda(D) \).  
For example, in many situations involving simple hypotheses, the Neyman-Pearson Lemma implies that the likelihood ratio \( \Lambda(D) = f_1(D)/f_0(D) \) defines the MP test statistic, assuming \( f_0(x) > 0 \) and \( f_1(x) > 0 \) for all \( x \in \mathbb{X} \).  
In the multivariate case, \( f_0 \) and \( f_1 \) are interpreted as joint densities with respect to a dominating measure \( \tau \); see \citet{LehmanAMS}.

A statistic \(A = A(D)\) is said to be \emph{ancillary} if its distribution is invariant under the hypotheses, i.e., \(f^A_0(u) = f^A_1(u)\) for all \(u\) in the support of \(A\). Thus, \(A\) cannot be used to discriminate between \(H_0\) and \(H_1\).

Let \(f_k^{T_1, T_2}(u,v)\) denote the joint density of two statistics \(T_1(D)\) and \(T_2(D)\) under \(H_k\), and let \(f_k^{T_1 \mid T_2}(u, v)\) denote the conditional density of \(T_1(D)\) given \(T_2(D) = v\) under \(H_k\). If \(T_1\) and \(T_2\) are independent under \(H_k\), then, for all \( u \in \mathrm{supp}(T_1) \) and \( v \in \mathrm{supp}(T_2) \), \(f_k^{T_1, T_2}(u,v) = f_k^{T_1}(u) \, f_k^{T_2}(v)\).

For conceptual simplicity, it will be assumed that in cases where a researcher plans to employ a test statistic \(S(D) = L(\Lambda(D))\), where \(L(u)\) is a strictly increasing  function and \(\Lambda(D)\) is MP, the test statistic in use is \(T(D) = W(S(D))\), with \(W = L^{-1}\) on the range of $L$. In such cases, we treat \(T(D)\) as the MP test statistic.

We now restate a 
strengthened version of a result from \citet{vexler2024characterization}, whose proof is revisited and streamlined in the present paper for completeness.

Let \(A = A(D)\) be a statistic satisfying \(f_1^A = f_0^A\); that is, \(A\) is ancillary. Suppose the test statistic \(T = T(D)\) can be written as a function of two components, \(T_N\) and \(A\), via \(T = \psi(T_N, A)\), where \(\psi\) is a  bivariate function. Assume further that \(T_N\) and \(A\) are independent under both \(H_0\) and \(H_1\). The following result holds.

\begin{proposition}\label{prop1}
	Let \(T_N \) be a statistic such that  there exists a strictly increasing  function  \( L(u)>0\)  for which \(f_1^{L(T_N)}(u) = u  f_0^{L(T_N)}(u)\), for all $u$. Then the test that rejects $H_0$ for large values of \(T_N\)  dominates the test based on \(T\) in terms of power.	
\end{proposition}
\renewcommand{\qedsymbol}{} 
\begin{proof}
	See Appendix.
\end{proof}
\renewcommand{\qedsymbol}{$\square$} 
This proposition illustrates that modifying a test statistic to be independent of an ancillary statistic can  improve the power of the original test.
\begin{remark}\label{rm00}
	\emph{
		In connection with the condition \(f_1^{L(T_N)}(u) = u  f_0^{L(T_N)}(u)\), for all $u$, in Proposition~\ref{prop1}, we observe the following.
		Assume we have a test statistic \( Y = Y(D) \), and the likelihood ratio \( L^Y(u) = f^Y_1(u)/f^Y_0(u) \) is a strictly increasing  function with an inverse function \( W(u) \). In this case, \( Y \) can be transformed to the form \( Y_N = L^Y(Y) \), and
		\[
		f^{Y_N}_k(u) = \frac{d}{du} {\Pr}_k\left\{ L^Y(Y) \le u \right\} = \frac{d}{du} {\Pr}_k\left\{ Y \le W(u) \right\} = f^Y_k\left(W(u)\right) \frac{d}{du} W(u), \quad k \in \{0, 1\}.
		\]
		This implies
		$	0<f^{Y_N}_1\left(u\right)/f^{Y_N}_0\left(u\right)\,=\,f_1^{Y}\left(W(u)\right)/f_0^{Y}\left(W(u)\right)
		\,=\,L^Y(W(u))
		\,=\,u.$
		\\
		Proposition~\ref{prop1} may alternatively be derived under the assumption that the identity 
		$  f^{L(T_N)}_1(u)$ $=$ $uf^{L(T_N)}_0(u) $ holds at $u=L(T_N)$ only (see the Appendix for details).
		\\
		For example,	according to \citet{sidak1999theory},  for location parameter testing problems, $ L^Y(Y)$ can be monotonic if and only if \( f^Y_0 \) and \( f^Y_1 \) are strongly unimodal \citep[pp. 32-33]{sidak1999theory}.
	}
\end{remark}

\section{Converse-type results on ancillarity and test optimality: examples and characterizations}\label{sc2}
In this section, Proposition~\ref{prop2} addresses the natural question arising from Proposition~\ref{prop1}: what constitutes an optimal choice of the ancillary statistic 
$A$ to ensure that the modified test statistic becomes most powerful (MP). Proposition~\ref{prop3} presents a result analogous to Basu's theorem, reformulated and proven within a decision-theoretic context. Proposition~\ref{prop4}  further illustrates how the proposed theoretical principles can be used to characterize the distribution of the  data. The proofs of the propositions are provided in the Appendix. Examples~\ref{ex1}-\ref{ex3} involving tests for location and scale are provided to illustrate the theoretical framework. 
Additional insights and implications of the theoretical results are discussed in Remark~\ref{r01} and in the Supplementary Material (see Remarks S1.1 and S1.2).

Assume that \( V = V(D) \) denotes a set of ancillary statistics with a joint density satisfying  
\( f^{V}_1(v) = f^{V}_0(v) \).

\begin{proposition}\label{prop2}  
	Let the mapping \( D \mapsto \left(T_N, V\right) \) be injective with a measurable inverse.  
	Assume \( T_N  \) is a test statistic that is independent of $V$ under both $H_0$ 
	and $H_1$, and that there exists a strictly monotonic function \( L(u)>0\) such that	\(	f^{L(T_N)}_1(u) = u\, f^{L(T_N)}_0(u).
	\)	Then \( T_N \) is MP.
\end{proposition}
Informally, Proposition~\ref{prop2} states that if a test statistic \( T_N \) is independent of a set of ancillary statistics \( V \), and the pair \( (T_N, V) \) suffices to reconstruct the underlying data, then the test based on \( T_N \) is MP.

\begin{remark}\label{r001}
	\emph{
		For vector-valued data \(D\), a natural extension of Proposition~\ref{prop2} can be used as follows.  
		Let \(T_N\) be conducted using a collection of statistics \(M\in \mathbb{R}^p\), $p\ge 1$, such that the mapping \(D \mapsto (M, V)\) is injective with a measurable inverse.  
		Assume that \(M\) is independent of \(V\) under both \(H_0\) and \(H_1\), and that there exists a strictly monotonic function \(L(u)>0\) satisfying
		\(
		f^{M}_1(m) = L(T_N(m))\, f^{M}_0(m), m\in \mathbb{R}^p.
		\)
		Then the test based on \(T_N\) is MP.
		\\		
		\noindent In this formulation, \(T_N\) is univariate, whereas \(M\) may be vector-valued. The  proof of this remark is given in the Appendix.
	}
\end{remark}

\begin{remark}\label{r01}
	\emph{
		Proposition~\ref{prop2} is plausibly related in spirit to Lemma 1 in \citet{kagan2002sufficiency}, though formulated, proved, and employed here in a decision-theoretic context.
	}
\end{remark}

An auxiliary result concerning data transformations and ancillary statistics is provided in the supplementary material (see Remark~S1.1).

In the next simple examples we illustrate how the conclusion of Proposition~\ref{prop2} may fail when one of
its structural conditions is violated.
\begin{example}[Failure of $T_N$ - ancillary independence $\Rightarrow$ $T_N$ is not MP.]
	\label{ex:dependence_notMP}
	Let $X_1,X_2$ be i.i.d. $N(\mu,1)$, and consider the simple one - sided problem:
$
	H_0:\mu=0$  versus  $H_1:\mu=\delta,\ \delta>0.$
	
	Define the ancillary statistic $V=X_1-X_2$,	
	noting that $V\sim N(0,2)$ under both $H_0$ and $H_1$.
	Consider the test statistic $T_N=X_1.$	
	The mapping $D=(X_1,X_2)\mapsto (T_N,V)=(X_1,\,X_1-X_2)$ is injective with measurable inverse,
	since $X_2=T_N-V$. However, $T_N$ is not independent of $V$:
	$
	\mathrm{Cov}(T_N,V)=1\neq 0.
	$
	It is clear that, for the one-sided alternative $\delta>0$,  the Neyman--Pearson (hence MP) test is based on $(X_1+X_2)/2$.
	Thus, when the independence condition in Proposition~\ref{prop2} is violated, the statistic $T_N$ need not yield an MP procedure.
\end{example}
\begin{example}[The pair $(T_N,V)$ does not determine $D$, and $T_N$ is not MP.]
	\label{ex:cont_ancillary_noninjective_notMP}
	Let $D=(X_1,X_2)$, where $X_1$ and $X_2$ are independent. Consider the simple hypotheses:
	$
	H_0:\ X_1\sim N(1,1),\ X_2\sim N(-1,1)$ versus 
	$
	H_1:\ X_1\sim N(-1,1),\ X_2\sim N(1,1)
	$.
	Define
$T_N=X_1+X_2,$ 	$V=|X_1-X_2|.$

	Note that, $V$ is ancillary in the two-regime sense, since, under $H_0$, one has $X_1-X_2\sim N(2,2)$, while under
	$H_1$ one has $X_1-X_2\sim N(-2,2)$. 
	
	The test statistic $T_N$ is independent of $V$ under both regimes. Indeed, 
	$
	\mathrm{Cov}(X_1+X_2,\ X_1-X_2)=\mathrm{Var}(X_1)-\mathrm{Var}(X_2)=0,
	$
	so $X_1+X_2$ and $X_1-X_2$ are independent; hence $X_1+X_2$ is independent of $|X_1-X_2|$.
	
	However, the mapping $D\mapsto (T_N,V)$ is not one-to-one: for fixed $t\in\mathbb{R}$ and $v>0$, the constraints
	$X_1+X_2=t$ and $|X_1-X_2|=v$ admit both cases $X_1-X_2=v$ and $X_1-X_2=-v$, which correspond to
	distinct values of $(X_1,X_2)$ while producing the same $(T_N,V)=(t,v)$.
	
	Finally, under both $H_0$ and $H_1$ one has $T_N=X_1+X_2\sim N(0,2)$, so $T_N$ cannot be MP. In contrast, the MP test statistic is the log-likelihood ratio $2(X_2-X_1)$.
\end{example}
\begin{example}[The third requirement in Proposition~\ref{prop2} fails although $T_N$ is independent of $V$;  $T_N$ is not MP.]
	\label{ex:nonmonotoneLR_notMP}
	Let $D=(X,Y)\in\mathbb{R}^2$, and assume that $X$ and $Y$ are independent under both $H_0$ and $H_1$.
	Assume that $Y\sim N(0,1)$ under both regimes, and define the ancillary statistic $V=Y$.
	Let $T_N=X$.
	The mapping $D\mapsto (T_N,V)=(X,Y)$ is injective with measurable inverse.
	
	Consider the simple hypotheses
	$
	H_0:\quad X \sim 0.5 N(2,1)+0.5 N(-2,1)$ versus
	$
	H_1:\quad X \sim N(0,1),
	$
	and write $\phi(\cdot)$ for the $N(0,1)$ density.
	Then $f_0^X(x)=0.5\phi(x-2)+0.5\phi(x+2)$ and $f_1^X(x)=\phi(x)$, and we have the ratio 
		\[
	\frac{f_1^X(X)}{f_0^X(X)}
	=\frac{2\phi(X)}{\phi(X-2)+\phi(X+2)}
	=\frac{e^{2}}{\cosh(2X)}.
	\]
	This ratio is not a monotone function of $X$ (it is even and is maximized at $X=0$).
	Therefore, the third requirement in Proposition~\ref{prop2} is not satisfied.
	
	The Neyman--Pearson (hence MP) test rejects $H_0$ for large values of the likelihood ratio $\Lambda(D)$.
	The rule ``reject $H_0$ for large values of $T_N=X$'' is not MP in this example,
	even though $T_N$ is independent of $V$ and $V$ is ancillary.
\end{example}

To state a result in the spirit of Basu's theorem, we formulate the following proposition.
\begin{proposition}\label{prop3}
	Assume the test statistic \( T_N \) is complete under \( H_k \), for \( k = 0 \) and/or \( k = 1 \).  
	If \( T_N \) is MP, then \( T_N \) is independent of \( V \) under both \( H_0 \) and \( H_1 \).
\end{proposition}

 The completeness assumption above is used in the same way as in Basu-type arguments
 (see Basu’s theorem and related discussion in  \cite{basu},  and also \cite{lehmann2006theory}).
 Completeness ensures that a zero-mean function of $T_N$ must be zero almost surely.
 Therefore, once one obtains an identity showing that the conditional law of $V$ given $T_N$
 differs from the marginal law only through such a zero-mean function of $T_N$,
 completeness forces this difference to vanish.
 Equivalently, the conditional distribution of $V$ given $T_N$ coincides with the marginal
 distribution of $V$, and hence $V$ and $T_N$ are independent.
A complete proof with full details is provided in the Appendix.
In Proposition~\ref{prop3}, completeness is invoked precisely to justify this converse step.

\begin{example}[Tests of the location in symmetric models]\label{ex1}
	{\em
		Assume we observe i.i.d.\ data points $X_1,X_2,\ldots,X_n$ that provide $D=\left\{X_1,\ldots,X_n\right\}$. 
		Let $X_1$ have a symmetric distribution with characteristic function $\phi_X(t)\ne 0$, for all $t\in \mathbb{R}^1$. Define $\bar{X}_n=\sum_{i=1}^nX_i/n$, and $\sigma^2=\text{var}(X_1)$. 		
		We consider the problem of testing $H_0:$ $\mu=0$ against $H_1:$ $\mu> 0$, where $\mu$ is the center of symmetry such that $f^{X_1}(\mu+u)=f^{X_1}(\mu-u)$ for every $u\in \mathbb{R}^1$, and where $f^{X_1}$ denotes the unimodal density function of $X_1$. A commonly used test statistic in this setting is $T=n^{0.5}\bar{X}_n/\sigma$.
		
		According to~\cite{behboodian1990some} and \cite{hamedani2003characterization}, $X_1,\ldots,X_n$ are symmetric if and only if $X_j-\bar{X}_n$ is symmetric for some $j\in \{1,\ldots,n\}$.		
		Given $X_j-\bar{X}_n$, $j\in \{1,\ldots,n\}$, and $T$, we can reconstruct $D$, since $X_j=X_j-\bar{X}_n+\sigma T/n^{0.5}$, $j\in \{1,\ldots,n\}$. 
		
		Define a linear combination $Z_n=a_1(X_1-\bar{X}_n)+\ldots+a_n(X_n-\bar{X}_n)$, where $a_k \in \mathbb{R}^1$, $k \in \{1,\ldots, n\}$. We have $Z_n=\sum_{i=1}^n X_i\left(a_i-\sum_{j=1}^na_j/n\right)$, and thus 
		$\mathrm{cov}\left(\bar{X}_n,Z_n\right)=\mathrm{E}\sum_{i=1}^n (X_i-\mu)^2\left(a_i-\sum_{j=1}^na_j/n\right)=0$.
		
		By virtue of the Cram\'{e}r--Wold theorem (i.e., since any linear combination of the components of the vector $\mathbf{V}=\left[X_1-\bar{X}_n,\ldots,X_n-\bar{X}_n\right]^\top$ is uncorrelated with $T$), it follows that $T$ and 
		$\mathbf{V}$ are independent when $X_1$ is normally distributed, where the operator \( ^\top \) denotes transposition. The ratio \( L(u) = f_1^T(u)/f_0^T(u) \) is exponential in \( u \), due to the normality of \( T \) under both hypotheses. This satisfies the condition \( f_1^{L(T)}(u) = u f_0^{L(T)}(u) \) required by Proposition~\ref{prop2}.   For a technical illustration of this principle, see Example~\ref{ex2}.
		Thus, the conditions of Proposition~\ref{prop2} are satisfied.
		
		We can also note that if $X_1$ is normally distributed and $\sigma$ is known, then $T$ is a complete statistic \citep[see Theorem 6.22]{lehmann2006theory}.	
	}
\end{example} 
Building on Example~\ref{ex1}, we present a result that can characterize the normal distribution. Suppose the vector $\mathbf{V}$ is ancillary with respect to the location parameter $\mu$; see, for example, \citet[p.~160]{welsh2011aspects} for discussion. This condition can also be  treated in terms of the  characteristic function of $\mathbf{V}$, denoted by
\begin{eqnarray*}
	\phi_{\mathbf{V}}\left(t_1,\ldots,t_n\right)&=&\mathrm{E}\exp\left\{\mathrm{i}\sum_{j=1}^n t_j\left(X_j - \bar{X}_n\right)\right\} 
	\,=\, \mathrm{E}\exp\left\{\mathrm{i} \sum_{j=1}^n t_j X_j - \mathrm{i} \sum_{j=1}^n t_j \sum_{k=1}^n X_k/n \right\} \\
	&=&  \mathrm{E}\exp\left\{\mathrm{i} \sum_{j=1}^n X_j \left(t_j - \sum_{k=1}^n t_k/n\right)\right\} 
	\,=\, \prod_{j=1}^n \phi_X\left(t_j - \sum_{k=1}^n t_k/n\right),
\end{eqnarray*}
where $\mathrm{i}^2 = -1$, $t_k \in \mathbb{R}^1$ for $k \in \{1, \ldots, n\}$, and $\phi_X$,  the characteristic function of $X_1$, is assumed to satisfy $\phi_X(\mu + t) = \phi_X(\mu - t)$ for all $t \in \mathbb{R}^1$.
\begin{proposition}\label{prop4}
	Let $X_1, X_2, \ldots, X_n$ be i.i.d.\ random variables such that $X_1$ has a symmetric unimodal distribution about $\mu$ and satisfies $\mathrm{E}X_1^2 < \infty$. Suppose that the sample mean $\bar{X}_n$ is a complete statistic under  either $H_0$ or $H_1$. Then, $X_1$ follows a normal distribution if and only if  $T=n^{0.5}\bar{X}_n/\sigma$ is the MP test statistic for the location parameter.
\end{proposition}

It is worth noting that, in a broad class of models, statistics based on $\bar{X}_n$ may serve as MP test statistics. 
For instance, consider testing the null hypothesis $H_0$ that $X_1,\ldots,X_n$ are i.i.d.\ observations from a density function $f_0$, against the alternative $H_1:$ $X_1,\ldots,X_n$ are i.i.d.\ from $f_1(u) = \exp\left(\theta u + \gamma\right) f_0(u)$, where $\theta$ and $\gamma$ are unknown; see \citet[pp.~97--98]{VexlerB}.

An additional characterization of the distribution via the form of an MP test is presented in Example~\ref{ex2} (see also \citet{pfanzagl1968characterization} in this context).

For testing the hypotheses considered in Example~\ref{ex1}, the statistics commonly used in practice are $T = n^{0.5} \bar{X}_n / \sigma$ and the Wilcoxon signed rank test statistic.
Proposition~\ref{prop4} raises a concern regarding the comparative performance of these procedures: the $t$-test may be superior to the Wilcoxon test when the underlying distribution deviates from normality; see \citet{vexler2018t} for discussion in this context.
\citet{rosenblatt2013another} showed that the Wilcoxon signed rank test is more efficient than the $T$-based test under the alternative hypothesis $H_1\colon 0 < \mu \le 1$, where $X_1 \sim (1 - \mu) N(0, 1) + \mu N(\theta, \sigma^2)$ and the parameters $\theta$ and $\sigma$ are known.
Furthermore, \citet{shiraishi1986optimum} proved that the asymptotic power of the Wilcoxon signed rank test equals that of the MP test under the local alternative $H_1\colon X_1 \sim (1 - \mu) G(x) + \mu \{G(x)\}^2$, where $\mu = \delta / n^{0.5}$, $G(x)$ is a known logistic distribution, and $\delta$ is fixed. \citet{shiraishi1986optimum} also concluded that the numerical values of the asymptotic relative efficiency of the Wilcoxon test with respect to the $T$-based test show no  loss, even under general contaminated alternatives.
\begin{example}[Tests of the location in an asymmetric exponential model]\label{ex2}
	{\em
		Let $X_1, X_2, \ldots, X_n$ be a sample from the negative exponential distribution
		\(
		F_e(x) = \left[1 - \exp\left\{-(x - \mu)/\sigma\right\}\right] I(x > \mu),
		\)
		with location parameter $\mu$ and scale parameter $\sigma > 0$. Consider testing hypotheses concerning $\mu$. Let $X_{(1)} \le X_{(2)} \le \cdots \le X_{(n)}$ denote the order statistics based on $\{X_1, \ldots, X_n\}$.
		Define the vector $\mathbf{V} = [V_1, \ldots, V_{n-1}]^\top$, where $V_i = X_{(i+1)} - X_{(i)}$ for $i \in [1, \ldots, n-1]$. The vector $\mathbf{V}$ is ancillary \citep{buehler1982some}, since
		\begin{eqnarray*}
			\Pr\left\{\bigcap_{i=1}^{n-1} \left\{ V_i < v_i \right\} 
			\right\}& = &
			\int\limits_{x_1 < \ldots < x_n} 
			I\left\{ \bigcap_{i=1}^{n-1} \left\{x_{i+1} - x_i < v_i \right\} \right\}
			n! \prod_{i=1}^n f^{X_1}(x_i)\, dx_1 \ldots dx_n\\
			& = &
			\int\limits_{z_1 < \ldots < z_n} 
			I\left\{ \bigcap_{i=1}^{n-1} \left\{z_{i+1} - z_i < v_i \right\} \right\}
			n! \prod_{i=1}^n f^{X_1}(z_i+\mu)\, dz_1 \ldots dz_n,
		\end{eqnarray*}
		where $f^{X_1}$ is the density function of $X_1$.			
		Given $X_{(1)}$ and $\mathbf{V}$, the original sample can be reconstructed via the recursion:
		\(
		X_{(2)} = V_1 + X_{(1)}, \, X_{(3)} = V_2 + X_{(2)},\ldots, \, X_{(n)} = V_{n-1} + X_{(n-1)}.
		\)
		\citet{ferguson1967characterizing} established the independence between the statistic \( T = X_{(1)} \) and the vector \( \mathbf{V} \). Assume, for simplicity, that when testing the location parameter \( \mu \), we consider the null hypothesis \( H_0: \mu = \mu_0 \), and let \( \mu_1 > \mu_0 \) be a possible alternative value. The ratio
		\[
		L(u) =f^T_1(u)/f^T_0(u)
		= e^{(\mu_0 - \mu_1)/\sigma}\, h(u)^{n-1}, \quad 
		\text{for } u > \mu_1 > \mu_0,
		\]
		with
		\[
		h(u)= \frac{1 - e^{-(u - \mu_1)/\sigma}}{1 - e^{-(u - \mu_0)/\sigma}},\quad
		h'(u) = 
		\frac{e^{-(u - \mu_1)/\sigma} - e^{-(u - \mu_0)/\sigma}}
		{\left( 1 - e^{-(u - \mu_0)/\sigma} \right)^2} > 0.
		\]
		Thus, \( L(u) \) is strictly increasing. In a manner similar to Remark~\ref{rm00}, we may consider the monotonic transformation \( T_N = L(T) \) as a theoretical device, which satisfies \( f^{T_N}_1(u) = u f^{T_N}_0(u) \). 
		Then, we conclude that the conditions of Proposition~\ref{prop2} hold, and therefore \( T \) is an MP test statistic.

		Now suppose $X_1, \ldots, X_n$ are i.i.d.\ observations from a continuous distribution. Redefine the ancillary vector 
		\(
		\mathbf{V} = [X_{(2)} - X_{(1)}, X_{(3)} - X_{(1)}, \ldots, X_{(n)} - X_{(1)}]^\top.
		\)
		According to \citet{govindarajulu1966characterization},  $T$ and $\mathbf{V}$ are independent if and only if $X_1$ follows the negative exponential distribution $F_e(x)$. Thus, if $T = X_{(1)}$ is an MP test statistic for the location, then $X_1, \ldots, X_n$ must follow the negative exponential distribution. This  yields a characterization of the negative exponential distribution via the form of the MP test statistic in this decision-making problem.
	}
\end{example}
\begin{example}[Tests of the scale (rate) of a gamma distribution]\label{ex3}
	{\em
		Let observations $X_1,\ldots,X_n$ be i.i.d.\ from a gamma distribution with the scale parameter $\theta$ (the rate parameter equals $1/\theta$) and the shape parameter $\kappa$. The random variables $X_i/\bar{X}_n = (\theta Y_i)/(\theta Z) = Y_i/Z$, where $Y_i\sim \mathrm{Gamma}(\kappa,1)$ and $Z\sim \mathrm{Gamma}(n\kappa,1/n)$, $i\in [1,\ldots,n]$. The distribution of the vector $\mathbf{V} = [X_1/\bar{X}_n,\ldots,X_n/\bar{X}_n]^\top$ is Dirichlet with parameter vector $[\kappa,\ldots,\kappa]^\top$, and it does not depend on $\theta$. 
		According to~\cite{lin2022new}, the statistics $\bar{X}_n$ and $\mathbf{V}$ are independent.
		When \( T = \bar{X}_n \sim \mathrm{Gamma}(n\kappa, \theta/n) \), the analysis of the ratio \( L(u)=f^{T}_1(u)/f^{T}_0(u) \) is the same as in Example 2.1.  
		Therefore, in this setting, \( \bar{X}_n \) serves as an MP test statistic for the scale parameter \( \theta \).				 
	}
\end{example}
In relation to Proposition~\ref{prop2}, a discussion on using distance correlation to compare test statistics with respect to their dependence on ancillary vectors is provided in the Supplementary Material (see Remark~S1.2).

\section{Distribution-free applications}\label{sc3}
This section presents simple examples that demonstrate the proposed technique in a transparent and practically applicable manner.
The following analysis examines ways to adapt existing test mechanisms and to compare alternative decision-making procedures using the framework developed in Section~\ref{sc2}.

In this section we use ancillarity in an approximate sense (see
Remark~\ref{Remark2:aprAnc} for this interpretation). Accordingly, the ancillary
invariance should be understood as approximate in finite samples, while still
serving as a useful design principle for constructing and comparing the testing
procedures.

Section~\ref{sc32} explores a simple modification of classical tests for normality. We show that the modified tests are asymptotically twice as efficient as their original counterparts under symmetric alternatives. The example is clear and self-contained, and it highlights elements that may extend to broader contexts of goodness-of-fit testing.

Section~\ref{sc33} addresses a multivariate mean testing problem, comparing Hotelling’s procedure with a simple trace-normalized test within the ancillary-independence framework. The results show that the trace-normalized test can outperform Hotelling’s procedure under heavy-tailed distributions, while Hotelling’s test remains preferable under normality.

%

\subsection{A straightforward approach to improving goodness-of-fit tests}\label{sc32}
We consider the problem of testing for normality based on a sample \( X_1, \ldots, X_n \).
There is a body of literature on tests for normality against asymmetric alternatives  \citep{lin1980simple} and on tests for normality against symmetric alternatives \citep{spiegelhalter1977test}. We illustrate an application of the presented framework to tests for symmetric alternatives.
Let the underlying distribution of $X_i$, $i \in[1,\ldots,n]$, be denoted by $F(u)$, which is assumed to be unknown.

\paragraph{Ancillary representation.} According to Proposition~\ref{prop2}, the vector  

\noindent 
$
\mathbf{V_1}
\,=\,\left[F\left(|X_1|\right)\right.$ $-$  $\left.F\left(-|X_1|\right),\ldots,F\left(|X_n|\right)\right.$ $-$  $\left.F\left(-|X_n|\right)\right]^{\top}$  
serves as an example of ancillary statistics that can be particularly effective for constructing a test of normality, assuming $F(u)$ were known, since $\Pr\left(|X_1|\le u\right)=F(u)-F(-u)$.  
The vectors $\mathbf{V_1}$ and $[\mathrm{sign}(X_1),\ldots,\mathrm{sign}(X_n)]^{\top}$ together can be used to reconstruct the original data \(\left( X_1, \ldots, X_n\right) \).
Using the empirical estimator of $\Pr\left(|X_1|\le u\right)$ based on  
\( |X_1|, \dots, |X_n| \), we approximate $\mathbf{V}_1$ by $\mathbf{\hat{V}}=n^{-1}\left[R_1^{+},\ldots,R^{+}_n\right]^{\top}$,  
where $R^{+}_i$ denotes the rank of $|X_i|$, $i\in[1,\dots,n]$.  

Define the order statistics  \( |X|_{(1)}, \ldots, |X|_{(n)} \) based on  \( |X_1|, \dots, |X_n| \).  
We invoke the following result: for any  distribution $F$ symmetric about zero, the random vectors $\mathbf{\hat{V}}$, $[\mathrm{sign}(X_1),\ldots,$ $\mathrm{sign}(X_n)]^{\top}$, and  
\( [|X|_{(1)}, \ldots, |X|_{(n)}]^{\top} \) are mutually independent \citep[p. 40]{sidak1999theory}.

This independence, along with Proposition~\ref{prop2}, motivates the development of tests for normality against symmetric alternatives based on the observations  
$|X|_{(1)}, \ldots, |X|_{(n)}.$  

\paragraph{Modified procedures.} 
For example, the Anderson-Darling test (AD), the Cram\'{e}r-von Mises test (CvM), and the Kolmogorov-Smirnov test (KS) involve the empirical distribution function of  
\( X_1, \dots, X_n \).  We propose modifications to these tests  
based on the following transformation: define $Z_i=(X_i-\bar{X}_n)\left\{n^{-1}\sum_{i=1}^n\left(X_i-\bar{X}_n\right)^2\right\}^{-0.5}$, with $\bar{X}_n=\sum_{i=1}^nX_i/n$, and  
apply the AD, CvM, and KS test strategies based on \(|Z|_{(1)}, \ldots, |Z|_{(n)}\) for assessing the null hypothesis $\Pr(|Z|<u)=2\Phi(u)-1$, where  
$\Phi(u)=(2\pi)^{-0.5}\int_{-\infty}^{u}\exp(-t^2/2)dt$, resulting in the modified versions of the tests, denoted as mAD, mCvM, and mKS, respectively. 
Note that it is common for the classical AD, CvM, and KS tests for normality to be based on \( Z_1, \ldots, Z_n \), in order to control their type I error  rates when the expectation and variance of the observations \( X_1, \ldots, X_n \) are unknown.
Regarding the equivalence between the statements $Z\sim N(0,1)$ and
${\Pr}\{|Z|\le u\}=2\Phi(u)-1$ for all $u\ge 0$ under the assumption that $Z$ is symmetrically distributed,
we refer the reader to \cite{vexler2020univariate}.

\paragraph{Asymptotic relative efficiency.} The ancillarity-based framework considered in Section~\ref{sc2} guides us to employ absolute values of the observations when applying the AD and CvM tests for normality. This, in turn, motivates an analysis of their asymptotic efficiency in the sense of \citet{inglot2000vanishing}. It can be shown that the modified Anderson–Darling and modified Cramér–von Mises tests are asymptotically twice as efficient as their original counterparts under symmetric alternatives. The corresponding proof is given in Section~S2 of the supplementary material.

\paragraph{Numerical illustration.}
The suggested approach above uses an empirical approximation of the theoretical principle shown in Proposition~\ref{prop2}.  
Its efficiency should be evaluated thoroughly, for example, through extensive Monte Carlo simulations.  
In this regard, we note that our attempts to modify the Shapiro--Wilk test, for example, by incorporating the expectations of \( |Z|_{(i)}, i\ge 1, \) or employing L-estimators based on \( |Z|_{(1)}, \ldots, |Z|_{(n)} \) in the SW manner, did not result in noticeable improvements.
Motivated by empirical evaluations, the modification (mSW) of the SW test that we suggest is as follows:  consider the data $D=\big\{-|Z|_{(1)}, \ldots, -|Z|_{(n)}$, $|Z|_{(1)}, \ldots, |Z|_{(n)}\big\}$,  
and compute the classic SW test statistic based on $D$. Under 
$H_0$, the mSW statistic is distributed independently of the parameters of the null distribution of the observations. 

To evaluate the extent of power loss when the proposed modifications are applied to asymmetric alternatives, we include the Spiegelhalter test (S) for comparison, as it is specifically constructed for testing normality against symmetric alternatives \citep{spiegelhalter1977test}.

\setlength{\tabcolsep}{1.8pt} 
\begin{table}[htbp]
	\centering
	\caption{
		{\bf Monte Carlo power of normality tests under various alternatives at $\alpha = 0.05$.}		The mSW, mAD, mCvM, and mKS denote modified tests. Under symmetric alternatives, the modified tests are expected to outperform their classical counterparts. Under asymmetric settings, any potential loss in power of the modified tests may be compared to that of the classical S test, which is designed for testing normality against symmetric alternatives.
	}
	\small
	\begin{tabular}{lccccccccc}
		\textbf{Alternative} & \textbf{SW} & \textbf{mSW} & \textbf{AD} & \textbf{mAD} & \textbf{CvM} & \textbf{mCvM} & \textbf{KS} & \textbf{mKS} & \textbf{S} \\
		$X\sim N(1,3^2)$ &0.051 & 0.047 & 0.049 & 0.050 & 0.048 & 0.050 & 0.048 & 0.048 & 0.048 \\
		$X\sim\mathrm{Exp(1)}$&0.968 & 0.514 & 0.936 & 0.334 & 0.896 & 0.348 & 0.837 & 0.394 & 0.341 \\
		$X=y_1-y_2,\, y_k\sim \mathrm{Exp(1)}$& 0.358 & 0.358 & 0.371 & 0.381 & 0.356 & 0.386 & 0.268 & 0.349 & 0.437 \\
		$X\sim \mathrm{Beta(0.5,0.5)}$
		&0.943 & 0.944 & 0.858 & 0.856 & 0.741 & 0.816 & 0.475 & 0.723 & 0.997 \\
		$X=y_1-y_2,\, y_k\sim \mathrm{Beta(0.5,0.5)}$
		&0.056 & 0.182 & 0.052 & 0.083 & 0.044 & 0.078 & 0.047 & 0.078 & 0.156 \\	 
		$X\sim \mathrm{Beta(2,5)}$
		&0.275 & 0.09 & 0.225 & 0.067 & 0.204 & 0.075 & 0.242 & 0.066 & 0.088 \\
		$X=y_1-y_2,\, y_k\sim \mathrm{Beta(2,5)}$
		&0.043 & 0.043 & 0.047 & 0.049 & 0.048 & 0.047 & 0.053 & 0.055 & 0.048 \\
		$X\sim \mathrm{Beta(2,2)}$ 
		&0.076 & 0.198 & 0.076 & 0.100 & 0.069 & 0.124 & 0.061 & 0.099 & 0.145 \\
		$X=y_1-y_2,\, y_k\sim \mathrm{Beta(2,2)},\,\,n=30$ 
		&0.036 & 0.061 & 0.043 & 0.048 & 0.042 & 0.051& 0.045 & 0.049 & 0.044 \\
		$X=y_1-y_2,\, y_k\sim \mathrm{Beta(2,2)},\,\,n=50$ 
		&0.038 & 0.073 & 0.047 & 0.055 & 0.048 & 0.059 & 0.047 & 0.054 & 0.036 \\
		$X=y_1-y_2,\, y_k\sim \mathrm{Beta(2,2)},\,\,n=150$ 
		&0.071 & 0.179 & 0.07 & 0.09 & 0.067 & 0.096 & 0.06 & 0.085 & 0.005 \\
		$X\sim \mathrm{Gamma(3,1)}$ 
		&0.562 & 0.202 & 0.478 & 0.106 & 0.434 & 0.107 & 0.433 & 0.093 & 0.147 \\
		$X=y_1-y_2,\, y_k\sim \mathrm{Gamma(3,1)}$
		&0.131 & 0.141 & 0.118 & 0.115 & 0.107 & 0.121 & 0.085 & 0.111 & 0.138 \\
		$X\sim \chi^2_{1}$
		&0.999 & 0.828 & 0.999 & 0.741 & 0.996 & 0.777 & 0.985 & 0.851 & 0.558 \\	
		$X=y_1-y_2,\, y_k\sim \chi^2_{1}$
		&0.665 & 0.699 & 0.729 & 0.753 & 0.736 & 0.737 & 0.617 & 0.741 & 0.766 \\
		$X\sim \mathrm{Triangular(-1,1)},\,\, n=30$
		&0.035 & 0.079 & 0.041 & 0.051 & 0.039 & 0.052 & 0.042 & 0.049 & 0.063 \\
		$X\sim \mathrm{Triangular(-1,1)},\,\, n=150$
		&0.214 & 0.538 & 0.109 & 0.134 & 0.077 & 0.132 & 0.057 & 0.107 & 0.021 \\
		$X\sim 0.5N(-1,1)+0.5N(1,1),\, n=30$
		&0.051 & 0.098 & 0.057 & 0.079 & 0.062 & 0.086 & 0.057 & 0.077 & 0.063 \\
		$X\sim 0.5N(-1,1)+0.5N(1,1),\, n=100$
		&0.110 & 0.227 & 0.131 & 0.241 & 0.134 & 0.255 & 0.101 & 0.178 & 0.016 \\
		$X=\mathrm{U[-1,1]}+ N(0,0.5^2),\, n=30$
		&0.042 & 0.074 & 0.048 & 0.054 & 0.051 & 0.059 & 0.051 & 0.059 & 0.050 \\
		$X=\mathrm{U[-1,1]}+ N(0,0.5^2),\,\, n=50$
		&0.045 & 0.084 & 0.052 & 0.064 & 0.054 & 0.072 & 0.051 & 0.068 & 0.038 \\
		$X=\mathrm{U[-1,1]}+ N(0,0.5^2),\,\, n=150$
		&0.087 & 0.184 & 0.100 & 0.137 & 0.092 & 0.151 & 0.077 & 0.13 & 0.003 \\
	\end{tabular}
	\label{tab:power-tests}
\end{table}
Under $H_0$, the considered test statistics have null distributions that are free of the unknown parameters.
Hence, we obtain their critical values by Monte Carlo simulation under $X_i \sim N(0,1)$, $i\ge 1$.
Specifically, for each value of $n$, the critical values for the SW, mSW, AD, mAD, CvM, mCvM, and S statistics
were computed from their empirical sampling distributions based on $50{,}000$ standard normal samples.

Table~\ref{tab:power-tests} summarizes selected Monte Carlo power results for various alternatives at the 5\% significance level, based on 50{,}000 replications per setting. Sample sizes are indicated in the alternative labels when different from $n=30$. The table highlights that  the modified tests tend to outperform their classical counterparts under symmetric alternatives, while remaining relatively competitive under asymmetric settings.
We note, however, that under strongly skewed alternatives the ancillary-guided construction may reduce sensitivity to the very features driving the departure from normality. This loss of power is of the same order as that of the classical 
test  S and is most visible for highly asymmetric laws (e.g., $\mathrm{Exp}(1)$). In such regimes the guiding ancillary structure is far from being approximately valid.
For example, in the case where the observations follow a $\mathrm{Beta(2,2)}$ distribution, the relative improvements in power were:  
$160.5\%$, $31.2\%$, $79.7\%$, and $62.2\%$, corresponding to the SW, AD, CvM, and KS test mechanisms, respectively. In several cases, we observed that the modification led to situations in which the original counterparts were biased, whereas their transformed versions consistently achieved power exceeding 5\%. For instance, consider the scenario where $X = y_1 - y_2$, with $y_k \sim \mathrm{Beta}(0.5, 0.5)$.

In a few cases, the classical S test outperforms the modified tests. However, there are  scenarios where the S test seems to be inconsistent. For example, Table~\ref{tab:power-tests} demonstrates such situations when \( n = 50, 100, 150 \) are considered, and the observations come from the mixture of normal distributions \( N(-1,1) \) and \( N(1,1) \). In these cases, the proposed tests display robust power characteristics.

Table~\ref{tab:power-tests} indicates that the proposed ancillary-guided modifications primarily act as symmetry-focused power concentrators. For symmetric alternatives, the modified tests often deliver substantial power gains over their classical counterparts, reflecting reduced ancillary leakage and increased sensitivity to the departure features targeted by the construction.

For every test considered, the nominal $\alpha=0.05$ lies within the $95\%$ confidence interval
$
\hat{\alpha} \pm 1.96\left\{\hat{\alpha}(1-\hat{\alpha})/50,000\right\}^{0.5},
$
confirming that the tests maintain the intended significance level within expected sampling variation.

To provide a practical interpretation, we also report an ``$n$-equivalent'' comparison for representative symmetric alternatives. For a fixed target power level, we determine the smallest sample size needed by the classical test to match the power of the modified test at the reported $n$, thereby quantifying the effective sample-size savings implied by Table~\ref{tab:power-tests}. For instance, under the symmetric Beta$(2,2)$ alternative with $n=30$, mSW achieves approximately $19.8\%$ power, whereas SW reaches $7.6\%$. To attain $19.8\%$ power, SW would require roughly $n\approx 80$ observations, corresponding to a reduction of about $50$ observations (approximately $63\%$ fewer) when using mSW.
Under the alternative $X\sim 0.5N(-1,1)+0.5N(1,1)$, at $n=100$ Table~\ref{tab:power-tests} reports experimental powers $0.131$ for AD and $0.241$ for mAD. Under the Monte Carlo design described in this section, we find that AD attains empirical power $0.241$ at about $n=180$. Thus, in this setting mAD at $n=100$ is roughly comparable to AD at about $n\approx 180$, corresponding to a sample-size reduction of about $44\%$.

Overall, the proposed modifications are simple to implement, and the experimental results confirm that the resulting procedures are effective in detecting deviations from normality under symmetry while maintaining robustness under broader alternatives.

\begin{remark}\label{nonp}
	{\em Our method requires identifying relevant ancillary statistics. In this context, we note that the concept of ancillarity can be treated through appropriate approximations \citep{skovgaard1985second}.  
		In many situations, maximum log-likelihood ratio type statistics may follow a $\chi^2$ asymptotic distribution independently of the underlying data distribution.  
		In nonparametric settings, the Wilcoxon test statistic based on the differences $Y_1=X_1-X_2,\,Y_2=X_3-X_4,\ldots$ can serve
		as an example of an ancillary statistic.}
\end{remark}

\subsection{Multivariate example: Hotelling and Trace-Normalized test statistics}\label{sc33}
\paragraph{Test statistics.} Let $\mathbf{X}_{1},\ldots,\mathbf{X}_{n}$ be i.i.d.\ $p$-dimensional random vectors with mean 
$\mu=\mathrm{E}(\mathbf{X}_{i})$ and covariance matrix $\Sigma=\mathrm{Cov}(\mathbf{X}_{i})$. 
We consider the one-sample problem: 
$
H_{0}:\ \mu=0 \quad \text{versus} \quad H_{1}:\ \mu\neq 0.
$
In this context, the classical Hotelling test statistic is
\[
T^{2}=n\,\bar{\mathbf{X}}^{\top} S^{-1}\bar{\mathbf{X}}, 
\qquad
\bar{\mathbf{X}}=n^{-1}\sum_{i=1}^{n}\mathbf{X}_{i},\quad
S=(n-1)^{-1}\sum_{i=1}^{n}(\mathbf{X}_{i}-\bar{\mathbf{X}})(\mathbf{X}_{i}-\bar{\mathbf{X}})^{\top}.
\]
An alternative test statistic, useful particularly when $p$ is relatively large, is the trace-normalized form
\[
T_{1}^{2}=\frac{n\,p\,\bar{\mathbf{X}}^{\top}\bar{\mathbf{X}}}{\mathrm{tr}(S)}.
\]
The Hotelling $T^{2}$ test is routinely used under multivariate normality of the observed data and also in broader non-normal 
settings. When inversion of $S$ is unstable or computationally 
burdensome, the $T_{1}^{2}$-based procedure provides a practical relaxation 
by replacing $S^{-1}$ with the scalar adjustment $\mathrm{tr}(S)^{-1}$; see \citet{FujikoshiHimenoWakaki2004}.

\paragraph{Ancillary structure.} In a nonparametric setting, and within the guidance of the theoretical framework of this paper, 
we ask under what conditions the use of $T^{2}$ or $T_{1}^{2}$ can be recommended for fixed, but 
possibly large, $p$. To this end we define the residual matrix
$
V = [\,\mathbf{X}_{1}-\bar{\mathbf{X}}, \ldots, \mathbf{X}_{n}-\bar{\mathbf{X}}\,]^{\top} \in \mathbb{R}^{n \times p},
$
the Gram matrix $W = V V^{\top}$, and note 
$
S = (n-1)^{-1} V^{\top} V.
$
Let $H = I_{n} - n^{-1}\mathbf{1}\mathbf{1}^{\top}$ denote the centering matrix, so that $V = H X$ and 
$W = H X X^{\top} H$, where $X$ is the $n \times p$ data matrix with rows $\mathbf{X}_{i}^{\top}$. 
In this framework, the matrix $V$ of centered observations is assumed to be an ancillary statistic, 
that is, its distribution does not depend on the mean parameter $\mu$ under either hypothesis. 

As an illustrative model, suppose that
$
\mathbf{X}_{i} = \mu + \Gamma \mathbf{Z}_{i}, \, i=1,\ldots,n,
$
where $\Gamma$ is a fixed $p\times p$ matrix and the $\mathbf{Z}_{i}$ are i.i.d.\ random vectors with 
$\mathrm{E}(\mathbf{Z}_{i})=0$. It is assumed that $\Gamma$ and the distribution of  $\mathbf{Z}_{1}$ do not depend on $\mu$. 

\paragraph{Dependence measure.} To assess and compare the dependence of the statistics $T^{2}$ and $T_{1}^{2}$ on the ancillary 
structure, we employ the vector correlation (RV) coefficient introduced by \cite{Escoufier1973}. 
For a scalar $U$ and a random vector $\mathbf{Y}$,
\[
\mathrm{RV}(U,\mathbf{Y})=\frac{\|\mathrm{Cov}(U,\mathbf{Y})\|^2}{\mathrm{Var}(U)\,\|\mathrm{Var}(\mathbf{Y})\|} \in[0,1],
\]
 where $\|\cdot\|$ denotes the Euclidean norm for vectors and the Frobenius norm for matrices.
 Then $\mathrm{RV}(U,\mathbf{Y})=0$ when $U$ and $\mathbf{Y}$ are independent, and 
$\mathrm{RV}(U,\mathbf{Y})=1$ when they are perfectly linearly dependent. In what follows, we will use the RV 
coefficient to approximate and compare the dependence of $T^{2}$ and $T_{1}^{2}$ on the ancillary 
statistic $V$.

By Lemma~S3.1 in the Supplementary Material (Section~S3), and since $T^{2}$ and $T_{1}^{2}$ 
are symmetric in the sample indices, we obtain
\[
\mathrm{RV}\!\left(T^{2},\,V\right)=0
\quad\text{and}\quad
\mathrm{RV}\!\left(T_{1}^{2},\,V\right)=0 .
\]
Consequently, we proceed to the next (quadratic) order by analyzing
\[
\mathrm{RV}\!\left(T^{2},\,\operatorname{vec}(W)\right)
\quad\text{and}\quad
\mathrm{RV}\!\left(T_{1}^{2},\,\operatorname{vec}(W)\right),
\]
in order to capture the leading nonzero quadratic component of the dependence between the
test statistics and the ancillary structure encoded by $V$. Here, for any matrix
$A \in \mathbb{R}^{m\times r}$, $\operatorname{vec}(A) \in \mathbb{R}^{mr}$ denotes the column-stacked
vectorization of $A$. In particular, since $V \in \mathbb{R}^{n\times p}$, we have
$W=V V^{\top}\in\mathbb{R}^{n\times n}$ and $\operatorname{vec}(W) \in \mathbb{R}^{n^{2}}$.

To study the ratio
\[
R \;=\; \frac{RV\!\left(T_1^2,\mathrm{vec}(W)\right)}{RV\!\left(T^2,\mathrm{vec}(W)\right)},
\]
we define the sphericity index
$
\psi \;=\; p\,\mathrm{tr}(\Sigma^2)\{\mathrm{tr}(\Sigma)\}^{-2}.
$
By the Cauchy–Schwarz inequality, we have $\psi\ge 1$ with equality iff $\Sigma\propto I_p$. We also denote an elliptical kurtosis factor as follows. Assume the elliptical fourth-moment identity
\[
\mathrm{E}\!\left\{ (Y^\top BY)(Y^\top CY) \right\}
= 2\,\mathrm{tr}(B\Sigma C\Sigma) + (\theta-1)\,\mathrm{tr}(B\Sigma)\,\mathrm{tr}(C\Sigma),
\qquad Y=\mathbf{X}_1-\mu,
\]
holds for symmetric $p\times p$ matrices $B,C$. The factor $\theta\ge 1$ is called the \emph{elliptical kurtosis factor}.
It links directly to Mardia’s multivariate kurtosis 
$
\beta_{2,p}=\mathrm{E}\!\left[\{(X-\mu)^\top \Sigma^{-1}(X-\mu)\}^2\right],
$
via
$\theta=\beta_{2,p}/\{p(p+2)\}$ under the illustrative model mentioned above, e.g.,  \cite{Mardia1974}. 
For example: when $\mathbf{X}_1\sim\mathcal{N}_p(\mu,\Sigma)$, $\beta_{2,p}=p(p+2)$ so $\theta=1$; when  $\mathbf{X}_1\sim$   
multivariate $t_\nu(\mu,\Sigma)$ with $\nu>4$, we have
$\theta=(\nu-2)/(\nu-4)>1.$

\paragraph{Asymptotic behavior.} We obtain the asymptotic approximation of $R$ in the form
\[
R \;=\; \frac{\big(2\psi+(\theta-1)p\big)^2}{\psi\,\big(2+(\theta-1)p\big)^2}+o(1),\qquad n\to\infty,
\]
where the following regimes can be considered:
(i) when $p$ is fixed, the remainder in the ratio expansion is of order $O(n^{-1})$;  
(ii) when both $p,n \to \infty$ proportionally with $p/n \to c \in (0,1)$, the remainder is larger, of order $O(n^{-1/2})$. 

A corresponding full derivation is given in Section~S3 of the Supplementary Material.

Under Gaussianity, $R\approx\psi\ge 1$, reflecting the well-known optimality of the $T^2$ decision-making procedure in this case. With heavy tails ($\theta>1$), $(\theta-1)p$ dominates and $R$ can fall below $1$. This indicates that the trace-normalized $T_1^2$ may reduce ancillary dependence relative to $T^2$, and in such strongly heavy-tailed settings $T_1^2$ can be superior to $T^2$.

\paragraph{Numerical illustration.}
To provide a brief numerical study, we considered $n=25$ and $p=15$. 
In the first design, data were generated from $N_{p}(\mu,\Sigma)$ with 
$\mu=(0,0.9,0,\ldots,0)^{\top}$ under $H_{1}$ and $\Sigma$ given by 
$\Sigma_{ii}=1$ for all $i$ and $\Sigma_{11}=10$. In this case the 
dependence ratio was estimated as $\widehat{R}\approx 1.4>1$. 
A Monte Carlo experiment with $30{,}000$ replications and empirical 
critical values at the $5\%$ level yielded power estimates of about $0.41$ 
for Hotelling’s $T^{2}$ and $0.22$ for the trace–normalized statistic 
$T_{1}^{2}$.

As a contrasting heavy–tailed design, we drew data from a multivariate 
$t_{5}$ distribution with mean vector 
$\mu_{t}=(0.9,0,\ldots,0)^{\top}$ under $H_{1}$ and covariance matrix 
equal to the identity. Here the estimated ratio was 
$\widehat{R}\approx 0.7<1$, and the same simulation scheme as in the 
normal case produced powers of about $0.57$ for $T^{2}$ and $0.84$ for 
$T_{1}^{2}$.

These power differences have direct practical implications. Under normality, Hotelling’s $T^2$ is more powerful, confirming its optimality in the regime for which it was designed. By contrast, under heavy tails the trace-normalized $T_1^2$ can be substantially stronger at the same sample size. The observed gain—from $0.57$ to $0.84$—corresponds to an absolute increase of about $0.27$ in the probability of detecting the mean shift. In this heavy-tailed regime, Hotelling’s $T^2$ would require roughly $1.5$--$2$ times as many observations (with $n\in (40,\ldots,50)$) to match the power achieved by $T_1^2$ at $n=25$.

This brief numerical study agrees with the theoretical prediction: 
$T^{2}$ is preferable in the $R>1$ regime, whereas $T_{1}^{2}$ is 
advantageous when $R<1$.


\section{Conclusion}\label{conclud}
This paper has introduced a relativity-based perspective for constructing, comparing, and improving test statistics by exploiting approximate independence from ancillary structures.
%
%
Within a decision-theoretic framework, we have shown that reducing the dependence between a test statistic and a vector of ancillary statistics can yield MP procedures under suitable conditions. Both direct and converse results in the spirit of Basu's theorem were established, offering a structural characterization of MP test statistics through their independence from ancillary quantities.  We also demonstrated that certain forms of MP test statistics implicitly characterize the data distribution.


The practical utility of the proposed approach was illustrated through simple examples and two nonparametric applications. We developed ancillary-guided modifications of classical normality tests, demonstrating their superiority in terms of asymptotic relative efficiency and supported by simulation studies, which show consistent power improvements under symmetric alternatives. We examined a multivariate mean testing problem, comparing Hotelling’s and trace-normalized statistics, and clarified the conditions under which each procedure is preferable. These procedures are straightforward to implement and highlight the effectiveness of the ancillary-independence framework in practice.

Several directions merit further investigation. Extending the methodology to multiple testing, high-dimensional inference, regression and model selection settings may reveal broader connections between ancillarity and the efficiency of statistical decision-making procedures. Within the proposed framework, alternative dependence measures—such as distance correlation, mutual information, or maximal correlation—may provide new perspectives on test construction. Data-driven approaches for selecting ancillary-guided transformations—potentially involving resampling or empirical likelihood—offer promising opportunities for future work.

\section*{Acknowledgments}
The authors thank Professor Teresa Ledwina (Institute of Mathematics, 
Polish Academy of Sciences, Poland) and Professor Chris J. Lloyd 
(Melbourne Business School, Australia) for insightful and helpful 
discussions related to the results presented in this work.
The authors are grateful to the Editor and the two anonymous referees for their constructive suggestions, which led to a substantial extension and improvement of the results in this paper.


\section*{Appendix}
\subsection*{Proof of Proposition~\ref{prop1}}
\begin{proof}
	Consider the elementary inequality: for all real values \( B \) and \( C \),
	\[
	(B - C)\left\{ I(B \ge C) - \delta \right\} \ge 0,
	\]
	where \( I(\cdot) \) is the indicator function and \( \delta \in [0,1] \).
	Let \( \delta = I(T \ge C^T) \) denote the rejection rule for a test based on the statistic \( T \) at threshold \( C^T \); that is, we reject \( H_0 \) when \( \delta = 1 \). Taking \( B = L(T_N) \), we obtain:
	\[
	\mathrm{E}_0 \left\{ (L(T_N) - C) I(L(T_N) \ge C) \right\} \ge \mathrm{E}_0 \left\{ (L(T_N) - C) I(T \ge C^T) \right\}.
	\]
	Now fix thresholds \( C \) and \( C^T \) such that the type I error rates are equal: \( {\Pr}_0\left\{(L(T_N) \ge C\right\} = {\Pr}_0(T \ge C^T) \). It follows that:
	\[
	\mathrm{E}_0\left[ I(L(T_N) \ge C) L(T_N) \right] \ge \mathrm{E}_0\left[ I(T \ge C^T)L( T_N) \right].
	\]
	Since the densities satisfy the identity \( u f_0^{L(T_N)}(u) = f_1^{L(T_N)}(u) \), the left-hand side becomes:
	\begin{eqnarray*}
		\mathrm{E}_0 \left[ I(L(T_N) \ge C) L(T_N) \right] &= &\int I(u \ge C) u f_0^{L(T_N)}(u) \, du = \int I(u \ge C) f_1^{L(T_N)}(u) \, du 
		\\
		&=& {\Pr}_1\left\{L(T_N) \ge C\right\}.
	\end{eqnarray*}
	To compute the right-hand side, recall that \( T = \psi(T_N, A) \), and use the independence of \( T_N \) and \( A \) under \( H_0 \):
	\[
	\begin{aligned}
		\mathrm{E}_0 \left[ I(T \ge C^T) L(T_N) \right] &= \mathrm{E}_0 \left[ I(\psi(T_N, A) \ge C^T) L(T_N) \right] \\
		&= \iint I(\psi(W(u),a) \ge C^T) u f_0^{L(T_N), A}(u, a) \, du \, da \\
		&= \iint I(\psi(W(u),a) \ge C^T) u f_0^{L(T_N)}(u)\, f_0^A(a) \, du \, da \\
		&= \iint I(\psi(W(u),a) \ge C^T) f_1^{L(T_N)}(u)\, f_0^A(a) \, du \, da 
	\end{aligned}
	\]
	where \( W \) denotes the inverse of \( L \). Then, applying the fact that \( f_0^A = f_1^A \), we have
	\[
	\begin{aligned}
		\mathrm{E}_1\left[ I(T \ge C^T) L(T_N) \right] 
		&= \iint I(\psi(W(u),a) \ge C^T) f_1^{L(T_N)}(u)\, f_1^A(a) \, du \, da \\
		&= 
		\mathrm{E}_1  \int I(\psi(W(u),A) \ge C^T) f_1^{L(T_N)}(u)  \, du  \\
		&= 
		\mathrm{E}_1 \mathrm{E}_1  I(\psi(W(L(T_N)),A) \ge C^T)\,=\,\, 
		\mathrm{E}_1 \mathrm{E}_1  I(T \ge C^T) \\
		&= {\Pr}_1(T \ge C^T),
	\end{aligned}
	\]
	This completes the proof. 
\end{proof}

\subsection*{Proof related to Remark~\ref{rm00}}
In this setting, we follow the proof scheme of Proposition~\ref{prop1} and note that
\[
\begin{aligned}
	\mathrm{E}_0\left[ I(L(T_N) \ge C) L(T_N) \right]& = 
	\mathrm{E}_0\left[ I(L(T_N) \ge C) \frac{f_1^{L(T_N)}(L(T_N))}{f_0^{L(T_N)}(L(T_N))}  \right]
	\\
	&=\int I(u \ge C) \frac{f_1^{L(T_N)}(u)}{f_0^{L(T_N)}(u)} f_0^{L(T_N)}(u) \, du\,\, =\,\, {\Pr}_1\left\{L(T_N) \ge C\right\}.
\end{aligned}
\]

\subsection*{Proof of Proposition~\ref{prop2}}
\begin{proof}
	Since \( D \mapsto \left(T_N, V\right) \) is injective with a measurable inverse, and \( L \) is strictly monotonic,  it follows  that the mapping $D \mapsto (L(T_N),V)$ is injective with a measurable inverse. Thus, the MP test statistic can be written as \( \Lambda(D) = g(L(T_N), V) \), for some bounded real-valued function \( g \). (For example, \( \Lambda(D) \) can be the corresponding likelihood ratio.)
	
	Without loss of generality, let \( L(u) \) be strictly increasing, and \( Y = L(T_N) \). In a manner similar to the proof of Proposition~\ref{prop1}, we have
	\[
	\text{E}_{0} \left[ I\left(Y \ge C\right) Y \right] \,\ge\, \text{E}_{0} \left[ I\left(\Lambda \ge C^\Lambda\right) Y \right],
	\]
	where \( C \) and \( C^\Lambda \) are chosen such that
	${\Pr}_{0}(Y \ge C) = \Pr_{0}(\Lambda \ge C^\Lambda),$
	i.e., the type I error rates \( {\Pr}_{0}(T_N \ge C^{T_N}) = {\Pr}_{0}(\Lambda \ge C^\Lambda) \), where the test threshold \( C^{T_N} = W(C) \), and \( W \) denotes the inverse of \( L \).
	
	Since \( f^{Y}_1(u) = u f^{Y}_0(u) \),   \( Y \) and \( V \) are independent, and \( f^{V}_0 = f^{V}_1 \), it follows that
	\begin{eqnarray*}
		\text{E}_{0} \left[ I(Y \ge C) Y \right] &=&
		\int I(u \ge C)\, u f^{Y}_0(u)\, du 
		= \int I(u \ge C)\, f^{Y}_1(u)\, du \\
		&=& {\Pr}_{1}(Y \ge C)\, =\, {\Pr}_{1}(T_N \ge C^{T_N}),
	\end{eqnarray*}
	and
	\begin{eqnarray*}
		&&\text{E}_{0} \left[ I(\Lambda \ge C^\Lambda) Y \right]\quad =\quad
		\text{E}_{0} \left[ I(g(Y, V) \ge C^\Lambda) Y \right] \\
		&&= \int\!\!\int I(g(u, v) \ge C^\Lambda) u f^{Y,V}_0(u, v) du dv 
		\,=\, \int\!\!\int I(g(u, v) \ge C^\Lambda) u f^{Y}_0(u)\, f^{V}_0(v) du dv \\
		&&=\, \int\!\!\int I(g(u, v) \ge C^\Lambda)\, f^{Y}_1(u)\, f^{V}_0(v) du dv 
		\,=\, \int\!\!\int I(g(u, v) \ge C^\Lambda)\, f^{Y}_1(u)\, f^{V}_1(v) du dv \\
		&&=\text{E}_{1}\int I(g(u, V) \ge C^\Lambda)\, f^{Y}_1(u)\, du\,\,=\,\,\text{E}_{1}\text{E}_{1} I(g(Y, V) \ge C^\Lambda)
		\,\,=\,\,\text{E}_{1}\text{E}_{1} I(\Lambda\ge C^\Lambda)
		\\
		&&=\quad {\Pr}_{1}(\Lambda \ge C^\Lambda).
	\end{eqnarray*}
	This completes the proof. 
	
	Note that, in this proof we use the representation of an MP rule as $I\left(\Lambda \ge C^\Lambda\right)$. In general, however, the same argument applies if we instead write the MP rule abstractly as the indicator of rejection by some existing MP level-$\alpha$ test, without requiring an explicit likelihood ratio $\Lambda$.
\end{proof}

\subsection*{Proof of Remark~\ref{r001}}
\begin{proof}
	The proof is based on that of Proposition~\ref{prop2} shown above.
	
	We begin proving that 	\(
	f_1^{L(T_N)}(u) = u\, f_0^{L(T_N)}(u)\), for $u>0$, when the conditions of  Remark~\ref{r001} are satisfied.
	Let \(h : \mathbb{R} \to \mathbb{R}\) be any bounded measurable function.
	By the definition of the density \(f_k^{L(T_N)}\), we have, for \(k=0,1\),
	\[
	\int_{\mathbb{R}} h(u)\, f_k^{L(T_N)}(u)\, \mathrm{d}u
	=
	\text{E}_k\bigl[h\bigl(L(T_N(M))\bigr)\bigr]
	=
	\int_{\mathbb{R}^p} h\bigl(L(T_N(m))\bigr)\, f_k^M(m)\, \mathrm{d}\tau(m).
	\]
	Then, for \(k=1\), we obtain
	\[
	\int_{\mathbb{R}} h(u)\, f_1^{L(T_N)}(u)\, \mathrm{d}u
	=
	\int_{\mathbb{R}^p} h\bigl(L(T_N(m))\bigr)\, f_1^M(m)\, \mathrm{d}\tau(m)
	=
	\int_{\mathbb{R}^p} h\bigl(L(T_N(m))\bigr)\, L\bigl(T_N(m)\bigr)\, f_0^M(m)\, \mathrm{d}\tau(m).
	\]
	On the other hand, under \(H_0\),
	\[
	\int_{\mathbb{R}} h(u)\, u\, f_0^{L(T_N)}(u)\, \mathrm{d}u
	=
	\text{E}_0\bigl[L(T_N(M))\, h\bigl(L(T_N(M))\bigr)\bigr]
	=
	\int_{\mathbb{R}^p} L\bigl(T_N(m)\bigr)\, h\bigl(L(T_N(m))\bigr)\, f_0^M(m)\, \mathrm{d}\tau(m).
	\]
	Hence, for every bounded measurable \(h\),
	\[
	\int_{\mathbb{R}} h(u)\, f_1^{L(T_N)}(u)\, \mathrm{d}u
	=
	\int_{\mathbb{R}} h(u)\, u\, f_0^{L(T_N)}(u)\, \mathrm{d}u.
	\]
	Since this equality holds for all bounded measurable functions \(h\),
	it follows that
	\[
	f_1^{L(T_N)}(u) = u\, f_0^{L(T_N)}(u)
	\]
	for Lebesgue-almost all \(u>0\).
	Therefore,  for \( Y = L(T_N) \), we have
	$ \text{E}_{0} \left[ I(Y \ge C) Y \right] 
	= {\Pr}_{1}(Y \ge C)\, =\, {\Pr}_{1}(T_N \ge C^{T_N}).$
	
	Now, with respect to the modified statement of  Proposition~\ref{prop2}, for \( Y = L(T_N) \),
	we  reconsider
	\begin{eqnarray*}
		&&\text{E}_{0} \left[ I(\Lambda \ge C^\Lambda) Y \right]\quad =\quad
		\text{E}_{0} \left[ I(g(M, V) \ge C^\Lambda) L(T_N(M)) \right] \\
		&&=\, \int\!\!\int I(g(u, v) \ge C^\Lambda)\, L(T_N(u)) \, f^{M,V}_0(u, v)\, du\, dv \\
		&&=\, \int\!\!\int I(g(u, v) \ge C^\Lambda)\, L(T_N(u)) \, f^{M}_0(u)\, f^{V}_0(v)\, du\, dv \\
		&&=\, \int\!\!\int I(g(u, v) \ge C^\Lambda)\, L(T_N(u)) \, f^{M}_0(u)\, f^{V}_1(v)\, du\, dv 
		=\, \int\!\!\int  I(g(u, v) \ge C^\Lambda) \, f^{M}_1(u)\, f^{V}_1(v)\, du\, dv \\
		&&=\, \int\!\!\int I(g(u, v) \ge C^\Lambda)\,  f^{M,V}_1(u,v)\, du\, dv 
		=\text{E}_{1}I(g(M, V) \ge C^\Lambda)
		\quad\,=\,\quad {\Pr}_{1}(\Lambda \ge C^\Lambda).
	\end{eqnarray*}
	This completes the proof, in a manner analogous to the proof of Proposition~\ref{prop2}.
	
\end{proof}
\subsection*{Proof of Proposition~\ref{prop3}}
\begin{proof}
	Let \( T_N \) be complete under \( H_k \), for \( k = 0 \) or \( k = 1 \).  
	It is clear that 
	\begin{eqnarray*}
		&& f_k^{V}(v) = \int f_k^{V}(v) f_k^{T_N}(u)\, du, \quad
		f_k^{V}(v) = \int f_k^{V|T_N}(v, u)\, f_k^{T_N}(u)\, du.
	\end{eqnarray*}
	Thus,
	\begin{eqnarray*}
		&& 0 = \int \left\{ f_k^{V|T_N}(v, u) - f_k^{V}(v) \right\} f_k^{T_N}(u)\, du = \text{E}_{k}
		\left\{ f_k^{V|T_N}(v, T_N) - f_k^{V}(v) \right\},
	\end{eqnarray*}
	where \( f_0^{V} = f_1^{V} \).  
	Therefore, \( \left\{ f_k^{V|T_N}(v, u) - f_k^{V}(v) \right\} = 0 \) almost surely under \( H_k \), since \( T_N \) is complete.
	
	Since $T_N$ is MP, and according to Proposition 2.4 of \citet{vexler2024characterization}, we have  
	\( f_0^{V|T_N} = f_1^{V|T_N} \), which implies
	\begin{eqnarray*}
		0 = \int \left\{ f_k^{V|T_N}(v, u) - f_k^{V}(v) \right\} f_k^{T_N}(u)\, du = \text{E}_{k}
		\left\{ f_r^{V|T_N}(v, T_N) - f_r^{V}(v) \right\}, \, r \in \{0,1\},\, r \ne k,
	\end{eqnarray*}
	and hence  
	\( \left\{ f_r^{V|T_N}(v, u) - f_r^{V}(v) \right\} = 0 \), almost surely.
	
	The proof is complete.
\end{proof}
\subsection*{Proof of Proposition~\ref{prop4}}
\begin{proof}
	If $X_1$ is normally distributed, then $T$ is MP, as shown in Example~\ref{ex1}.
	
	Conversely, suppose $T$ is MP. By Proposition~\ref{prop3}, $T$ and $\mathbf{V}$ are independent. This independence implies that $\mathrm{E}\left( \bar{X}_n \mid \mathbf{V} \right) = \mu$ and  
	$\mathrm{var}\left( \bar{X}_n \mid \mathbf{V} \right) = \sigma^2/n$. 	
	By Theorem 7.2.1 of \citet{bryc1995normal}, this conditional structure implies that $X_1$ must be normally distributed.
\end{proof}

\bmhead{Supplementary information} 
This supplement presents technical proofs of results stated in the main text which are omitted there for brevity.	
				Section~S1 
			collects remarks that support or extend points made in the main article, including additional results.
			With respect to the results shown in Section~\ref{sc32}, Section~S2 provides a detailed analysis of the Anderson–Darling (AD) and Cramér–von Mises (CvM) tests applied to both the raw observations $X_1,\ldots,X_n$ and their absolute values. We examine the asymptotic relative efficiency (ARE) of these tests under a class of symmetric contiguous alternatives. The analysis, based on the framework of \citet{inglot2000vanishing}, demonstrates that applying the tests to the absolute values—i.e., folding the data—increases their asymptotic efficiency.		
			Section~S3 (“Supplementary to Section~\ref{sc33}”) presents Lemma~S3.1 and Proposition~S3.1, which develop the RV–based dependence analysis for Hotelling’s $T^2$ and the trace–normalized $T_1^2$ statistics defined in Section~\ref{sc33}.

	
	\bibliographystyle{imsart-nameyear} 
	\bibliography{testsApproachREF1}

\end{document}


\renewcommand{\theequation}{S.\arabic{equation}}
	\setcounter{equation}{0}
	\renewcommand\thesection{S\arabic{section}}
	\renewcommand\thesubsection{S\arabic{section}.\arabic{subsection}}
	
	\maketitle

	\section*{Overview}	
		
	This document provides supplementary material for the paper titled \emph{“A Relativity-Based Framework for Statistical Testing Guided by the Independence of Ancillary Statistics: Methodology and Nonparametric Illustrations.”} This supplement presents technical proofs of results stated in the main text which are omitted there for brevity.
	
	Section~\ref{S0} 
	collects remarks that support or extend points made in the main article, including additional results.
		With respect to the results shown in Section~\ref{sc32}, Section~\ref{Sn2} provides a detailed analysis of the Anderson–Darling (AD) and Cramér–von Mises (CvM) tests applied to both the raw observations $X_1,\ldots,X_n$ and their absolute values. We examine the asymptotic relative efficiency (ARE) of these tests under a class of symmetric contiguous alternatives. The analysis, based on the framework of \citet{inglot2000vanishing}, demonstrates that applying the tests to the absolute values—i.e., folding the data—increases their asymptotic efficiency.		
	Section~\ref{S1} (“Supplementary to Section~\ref{sc33}”) presents Lemma~\ref{SL1} and Proposition~\ref{thm:ratio}, which develop the RV–based dependence analysis for Hotelling’s $T^2$ and the trace–normalized $T_1^2$ statistics defined in Section~\ref{sc33}. 
	 
	 	All references cited in this supplement are listed at the end of the document.

	\section{\ Additional Remarks}\label{S0}
	\begin{remark}\label{Sr02}
		\emph{
			Assume that a test statistic \( Q \) and a set of ancillary statistics \( V \) can be constructed from modified data \( D^M \).  
			Suppose the mapping \( D^M \to (Q, V) \) is injective with a measurable inverse.  
			If \( Q \) is independent of \( V \) under \( H_k \), \( k \in \{0,1\} \), and \( f^{Q}_1(u) = u\, f^{Q}_0(u) \),  
			then \( Q \) is MP among all test statistics that can be constructed from \( D^M \) alone.  
			For example, in several settings, it is natural to take \( D^M \) as the set of ranks of the observations; see, e.g.,~\citet{sidak1999theory,shiraishi1986optimum}.
		}
	\end{remark}
\begin{remark}\label{2.4}
	{\em
		\citet{anderson2016note} discussed criteria for evaluating and comparing the performance of test statistics. Suppose, in a given problem, we consider two test statistics, say $T_1$ and $T_2$, along with two ancillary vectors $\mathbf{V_1}$ and $\mathbf{V_2}$. It is possible that $\mathbf{V_1} = \mathbf{V_2}$. The underlying data $D$ can be reconstructed either from $(T_1, \mathbf{V_1})$ or from $(T_2, \mathbf{V_2})$. When selecting between $T_1$ and $T_2$ for use in a study, the distance correlation introduced by \citet{szekely2007measuring} between a test statistic and its corresponding ancillary vector may serve as a criterion. This approach is applicable in both parametric and nonparametric comparisons of competing test procedures.
	}
\end{remark}

\section{\ Anderson-Darling and Cram\'{e}r-von Mises Tests Based on $X$ and  $|X|$: Formulation, Symmetry, and Asymptotic Relative Efficiency.}\label{Sn2}
We begin by analyzing the Anderson–Darling (AD) tests in detail. The corresponding Cramér–von Mises (CvM) tests can be studied analogously, and we omit their derivation for brevity. Our main reference throughout is \citet{inglot2000vanishing}.

\subsection{Testing problems and statistics}

Without loss of generality, we assume that  $X_1,\dots,X_n$ are i.i.d.\ from a distribution function $F$ with $\mathrm{E}\left(X_1\right)=0$ and $\mathrm{var}\left(X_1\right)=1$. We consider
\[
H_0:\;F=\Phi, \qquad\text{vs.}\qquad H_1:\;F\neq \Phi,
\]
where $\Phi$ is the standard normal cdf. Define $U_i=\Phi(X_i)\in(0,1)$ so that under $H_0$, $U_i\sim\mathrm{Unif}(0,1)$.

\paragraph{Anderson - Darling on $X$ (test $T$).}
Let $F_n^{U}$ denote  the empirical distribution function based on the sample $U_1,\ldots,U_n$. The AD statistic is
\[
T \;=\; n\int_0^1 \frac{\{F_n^{U}(t)-t\}^2}{t(1-t)}\,dt.
\]

\paragraph{Anderson--Darling on $|X|$ (test $T_1$).}
Let $Z_i=|X_i|$ and define $Y_i = 2\Phi(Z_i)-1\in(0,1)$; under $H_0$, $Y_i\sim\mathrm{Unif}(0,1)$. 
Let  $F_n^{Y}$ denote  the empirical distribution function based on the sample $Y_1,\ldots,Y_n$.
Define
\[
T_1 \;=\; n\int_0^1 \frac{\{F_n^{Y}(v)-v\}^2}{v(1-v)}\,dv.
\]

\subsection{Local alternatives on the uniform scale}
\label{ssec:alts}
Following \citet{inglot2000vanishing}, we evaluate the efficiency of the tests under the standard class of contiguous alternatives.
\paragraph{Test $T$:} On the $U$-scale we adopt the standard contiguous alternative defined
by densities (with respect to Lebesgue measure on $[0,1]$),
\[
p_\theta(u)\;=\;1+\theta\,a(u), \qquad u\in(0,1),\quad \theta\to 0,
\]
with bounded function $a(u)$, and normalization $\int_0^1 a(u)\,du=0$ and $\int_0^1 a(u)^2\,du=1$.

%

\begin{assumption}[Symmetry]
	\label{ass:symm}
	The alternatives are symmetric about zero on the \( X \)-scale; that is, the perturbation function \( a_X(x) \) satisfies \( a_X(x) = a_X(-x) \). This implies that on the \( U \)-scale, the score function \( a(u) \) satisfies
	\[
	a(u) = a(1 - u), \qquad u \in (0,1).
	\]
\end{assumption}

\begin{proof}[Justification of symmetry in \( a(u) \)]
	Let \( U = \Phi(X) \), so that \( 1 - U = \Phi(-X) \). Since \( X \overset{d}{=} -X \) under the symmetry assumption, it follows that \( U \overset{d}{=} 1 - U \). Therefore, the density satisfies \( p_\theta(u) = p_\theta(1 - u) \). Using the local alternative expansion \( p_\theta(u) = 1 + \theta a(u) \), we conclude \( a(u) = a(1 - u) \). \qedhere
\end{proof}

\paragraph{Test $T_1$:}

Under the alternative above, the density of \(X\) is \(f(x) = \phi(x) [1 + \theta a(\Phi(x))]\), where $\phi(x)=\exp(-x^2/2)/(2\pi)^{0.5}$. Since the alternative is symmetric, the density of \(|X|\) at \(y\) (for \(y \geq 0\)) is:
\[
g(y) = f(y) + f(-y) = \phi(y) [1 + \theta a(\Phi(y))] + \phi(-y) [1 + \theta a(\Phi(-y))].
\]
Using \(\phi(-y) = \phi(y)\) and \(\Phi(-y) = 1 - \Phi(y)\), and using  symmetry (\(a(1-\Phi(y)) = a(\Phi(y))\)), this simplifies to:
\[
g(y) = 2\phi(y) [1 + \theta a(\Phi(y))].
\]
Under the null, the density of \(|X|\) is \(g_0(y) = 2\phi(y)\).

Now, to find the density of \(Y\), we note that \(Y = 2\Phi(|X|) - 1\). The cdf of \(Y\) is:
\[
{\Pr}(Y \leq v) = {\Pr}\left(2\Phi(|X|) - 1 \leq v\right) = {\Pr}\left(|X| \leq \Phi^{-1}\left(\frac{v+1}{2}\right)\right).
\]
Let \(z = \Phi^{-1}\left(\frac{v+1}{2}\right)\). The density of \(Y\) is the derivative of the cdf:
\[
p^Y_\theta = g(z)  \left| \frac{dz}{dv} \right|.
\]
From \(z = \Phi^{-1}\left(\frac{v+1}{2}\right)\), we have \(\Phi(z) = \frac{v+1}{2}\). Differentiating both sides with respect to \(v\):
\[
\phi(z) \frac{dz}{dv} = \frac{1}{2} \implies \frac{dz}{dv} = \frac{1}{2\phi(z)}.
\]
Substituting \(g(z) = 2\phi(z) [1 + \theta a(\Phi(z))]\):
\[
p^Y_\theta(v) = 2\phi(z) [1 + \theta a(\Phi(z))] \cdot \frac{1}{2\phi(z)} = 1 + \theta a(\Phi(z)).
\]
Since \(\Phi(z) = \frac{v+1}{2}\), we get:
\[
p^Y_\theta(v) = 1 + \theta a\left( \frac{v+1}{2} \right).
\]
Thus, the alternative density for \(Y\) is 
\[p^Y_{\theta}(v) = 1 + \theta b(v),\] 
where \(b(v) = a\left( \frac{v+1}{2} \right)\).

\subsection{The MP test of $H_0$: $U\sim$ Unif$(0,1)$ (or equivalently, $Y\sim$ Unif$(0,1)$)
	against the simple alternative $U\sim p_{\theta}$ (or equivalently,  $Y\sim p^Y_{\theta}$)}
According to ~\citet[p. 224]{inglot2000vanishing}, the corresponding locally most powerful test is given by:
\[
T^{\ast}
= \bigl(\sqrt{n}\,\sigma_{0n}\bigr)^{-1}
\sum_{i=1}^n \bigl(\log p_{\theta_n}(X_i) - e_{0n}\bigr),
\qquad
e_{0n} = \operatorname{E}_{\theta_0}\!\left[\log p_{\theta_n}(X)\right],
\quad
\sigma_{0n}^2 = \operatorname{var}_{\theta_0}\!\left[\log p_{\theta_n}(X)\right]
\]
with $0<\theta_n\to 0$ as $n\to \infty$.

\subsection{Drift functionals}
\label{ssec:functionals}
As shown in \citet{inglot2000vanishing}, the Pitman slopes of the test statistics can be expressed via weighted integrals of the following score function.

\paragraph{Test $T$:} Define the integrated score on $(0,1)$
\[
A(t) \;=\; \int_0^t a(x)\,dx, \qquad 
A_w(t) \;=\; \frac{A(t)}{\sqrt{t(1-t)}}.
\]
\paragraph{Test $T_1$:} For the test based on absolute values, define
\[
B(v) \;=\; \int_0^v b(s)\,ds, \qquad 
B_w(v) \;=\; \frac{B(v)}{\sqrt{v(1-v)}}.
\]

Under symmetry we have the \emph{mapping identity}
\begin{eqnarray}\nonumber
	B(u)&=&\int_{0}^{u} a\left(\frac{x+1}{2}\right)dx=2\int_{0.5}^{(u+1)/2} a\left(x\right)dx	
	\\
	&=&2A\left(\frac{u+1}{2}\right),
	\label{Eq1}
\end{eqnarray}
since 
the condition 
\(\int_{0}^1a(x)dx=0\) and Assumption \ref{ass:symm} lead to 
\[\int_{0}^{0.5}a(x)dx+\int_{0.5}^{1}a(x)dx=0, \qquad 
\int_{0}^{0.5}a(x)dx+\int_{0.5}^{1}a(1-x)dx=0
\]
meaning that \(\int_{0.5}^1a(x)dx=0\).

\subsection{A symmetry reduction for $\|A_w\|^2$ and a comparison inequality for $\|B_w\|^2$}
Let 
\begin{eqnarray*}\nonumber
	\|A_w\|^2 \;=\; \int_0^1 \frac{A(t)^2}{t(1-t)}\,dt \qquad\mbox{and}\qquad
	\|B_w\|^2 \;=\; \int_0^1 \frac{B(t)^2}{t(1-t)}\,dt.
\end{eqnarray*}

We show that  $A(1-t)=-A(t)$.
\begin{eqnarray*}\nonumber
	A(1-u)&=&\int_{0}^{1-u} a\left(x\right)dx=\int_{0}^{1-u} a\left(1-x\right)dx=
	-\int_{1}^{u} a\left(x\right)dx=\int_{u}^{1} a\left(x\right)dx\\
	&=&-A(u),
	\label{Eq2}
\end{eqnarray*}
since 
the condition 
\(\int_{0}^1a(x)dx=0\) leads to 
\[\int_{0}^{u}a(x)dx+\int_{u}^{1}a(x)dx=0, \qquad 
\int_{0}^{u}a(x)dx+\int_{u}^{1}a(x)dx=0
\]
meaning that \(\int_{u}^1a(x)dx=-\int_{0}^ua(x)dx=-A(u)\).

Hence,
\begin{eqnarray}\nonumber
	\|A_w\|^2 \;=\; \int_0^1 \frac{A(t)^2}{t(1-t)}\,dt =
	\int_0^{0.5}\frac{A(t)^2}{t(1-t)}\,dt +\int_{0.5}^{1}\frac{A(t)^2}{t(1-t)}\,dt 
	\\
	=\int_0^{0.5}\frac{\left(A(1-t)\right)^2}{t(1-t)}\,dt +\int_{0.5}^{1}\frac{A(t)^2}{t(1-t)}\,dt 
	\;=\; 2\int_{0.5}^{1} \frac{A(t)^2}{t(1-t)}\,dt.
	\label{Eq3}
\end{eqnarray}

Thus, we conclude that
\begin{proposition}[Norm comparison]
	\label{thm:norm}
	Under Assumption~\ref{ass:symm} and for any nondegenerate $a$ (equivalently $A\not\equiv 0$),
	\[
	\|B_w\|^2 \;>\; 2\,\|A_w\|^2.
	\]
\end{proposition}
\begin{proof}
	We have that Equation~(\ref{Eq1}) gives	
	\[
	\|B_w\|^2 \;=\; \int_0^1 B_w(v)^2\,dv 
	\;=\; 4\int_{0}^{1}  \frac{A\left((t+1)/2\right)^2}{t(t-1)}\,dt\,=\,
	4\int_{0.5}^{1}  \frac{A\left(z\right)^2}{(1-z)(2z-1)}\,dz.
	\]
	For $z\in[0.5,1)$, the denominator $(1-z)(2z-1)=(1-z)(z+z-1)< (1-z)z$. Result~(\ref{Eq3})
	completes the proof.	
\end{proof}

\subsection{Asymptotic Relative Efficiency}
According to Theorem~5.2(d) of \citet{inglot2000vanishing}, we have
\[
\text{ARE}(T,T^{\ast}) = 2\|A_w\|^2 < \|B_w\|^2 = \text{ARE}(T_1,T^{\ast})/2,
\]
where $\text{ARE}$ denotes the asymptotic relative efficiency, defined as the approximate ratio of the sample sizes required by a given test and the most powerful (MP) test to attain the same power under a sequence of local alternatives $p_{\theta_n}$, with $\theta_n \to 0$ and $n \to \infty$, such that $\theta_n^\gamma \Phi^{-1}(1-\alpha_n)\to 0$ for some $\gamma \in (0,1)$ and $\alpha_n \to 0$; see \citet[p.~231]{inglot2000vanishing}.

Therefore, under symmetric alternatives around zero, the test \( T_1 \) based on the absolute values \( |X_1|, \ldots, |X_n| \) is asymptotically more efficient than the classical Anderson--Darling test \( T \), which is applied directly to the raw observations.

	\section{Supplementary to Section~\ref{sc33}}\label{S1}
\begin{lemma}\label{SL1}
	Let $\mathbf{X}_{1},\ldots,\mathbf{X}_{n}$ be i.i.d.\ $p$-dimensional random vectors with mean $\mu=\mathrm{E}(\mathbf{X}_{i})$ and covariance $\Sigma=\mathrm{var}(\mathbf{X}_{i})$. 
	Write $\bar{\mathbf{X}}=n^{-1}\sum_{i=1}^{n}\mathbf{X}_{i}$ and 
	$
	\mathbf{V}=\big(\mathbf{X}_{1}-\bar{\mathbf{X}},\ldots,\mathbf{X}_{n}-\bar{\mathbf{X}}\big)^{\!\top}\in\mathbb{R}^{n\times p}.
	$
	Let $T^{2}$ be Hotelling’s statistic (or, more generally, any statistic that is symmetric under permutations of the sample indices). Then, for each $i\in\{1,\ldots,n\}$,
	\[
	\mathrm{E}\!\left[T^{2}\,(\mathbf{X}_{i}-\bar{\mathbf{X}})\right]=\mathbf{0}\in\mathbb{R}^{p}.
	\]
	Consequently, for any $b\in\mathbb{R}^{n}$,
	\[
	\mathrm{E}\!\left[T^{2}\, b^{\top}\mathbf{V}\right]=0
	\quad\text{and}\quad
	\mathrm{Cov}\!\left(T^{2},\, b^{\top}\mathbf{V}\right)=0.
	\]
	The same conclusions hold with $T^{2}$ replaced by $T_{1}^{2}$.
\end{lemma}

\begin{proof}
	Since $T^{2}$ is a symmetric statistic of the sample $\{\mathbf{X}_{1},\ldots,\mathbf{X}_{n}\}$ and the observations are i.i.d., the vector of mixed moments $\mathrm{E}(T^{2}\mathbf{X}_{j})$ does not depend on $j$. Hence,
	\[
	\mathrm{E}\!\left[T^{2}\,\bar{\mathbf{X}}\right]
	=\mathrm{E}\!\left[T^{2}\,\frac{1}{n}\sum_{j=1}^{n}\mathbf{X}_{j}\right]
	=\frac{1}{n}\sum_{j=1}^{n}\mathrm{E}\!\left[T^{2}\mathbf{X}_{j}\right]
	=\mathrm{E}\!\left[T^{2}\mathbf{X}_{1}\right].
	\]
	Therefore, for each $i$,
	\[
	\mathrm{E}\!\left[T^{2}\,(\mathbf{X}_{i}-\bar{\mathbf{X}})\right]
	=\mathrm{E}\!\left[T^{2}\mathbf{X}_{i}\right]-\mathrm{E}\!\left[T^{2}\bar{\mathbf{X}}\right]
	=\mathrm{E}\!\left[T^{2}\mathbf{X}_{1}\right]-\mathrm{E}\!\left[T^{2}\mathbf{X}_{1}\right]
	=\mathbf{0}.
	\]
	For any $b\in\mathbb{R}^{n}$,
	\[
	\mathrm{E}\!\left[T^{2}\, b^{\top}\mathbf{V}\right]
	=\sum_{i=1}^{n} b_{i}\,\mathrm{E}\!\left[T^{2}(\mathbf{X}_{i}-\bar{\mathbf{X}})\right]
	=0.
	\]
	Since $\mathrm{E}(b^{\top}\mathbf{V})=0$ by construction, it follows that
	\[
	\mathrm{Cov}\!\left(T^{2},\, b^{\top}\mathbf{V}\right)
	=\mathrm{E}\!\left[T^{2}\, b^{\top}\mathbf{V}\right]-\mathrm{E}(T^{2})\,\mathrm{E}\!\left(b^{\top}\mathbf{V}\right)
	=0- \mathrm{E}(T^{2})\cdot 0
	=0.
	\]
	The argument for $T_{1}^{2}$ is identical, since it is also symmetric in the sample indices.
\end{proof}

\begin{proposition} [The asymptotic approximation of $R$]
	\label{thm:ratio}
	Let $\mathbf{X}_{1},\ldots,\mathbf{X}_{n}$ be i.i.d.\ $p$-dimensional random vectors with 
	$\mathrm{E}(\mathbf{X}_{i})=\mu$, $\mathrm{Cov}(\mathbf{X}_{i})=\Sigma$, and finite fourth moments. 
	Assume the elliptical fourth-moment identity
	\begin{equation}
		\mathrm{E}\!\left\{ (Y^\top B Y)(Y^\top C Y) \right\}
		= 2\,\mathrm{tr}(B\Sigma C\Sigma) + (\theta-1)\,\mathrm{tr}(B\Sigma)\,\mathrm{tr}(C\Sigma),
		\qquad Y=\mathbf{X}_1-\mu,
		\label{eq:elliptical-identity}
	\end{equation}
	for symmetric $p\times p$ matrices $B,C$, where $\theta\ge 1$ is the elliptical kurtosis factor. 
	Let
	\[
	T^{2}=n\,\bar{\mathbf{X}}^{\top} S^{-1}\bar{\mathbf{X}}, 
	\qquad 
	T_{1}^{2}=\frac{n p\,\bar{\mathbf{X}}^{\top}\bar{\mathbf{X}}}{\mathrm{tr}(S)},
	\qquad
	\psi=\frac{p\,\mathrm{tr}(\Sigma^2)}{\{\mathrm{tr}(\Sigma)\}^{2}}\ge 1,
	\]
	and denote $W=VV^\top$, where $V=[\,\mathbf{X}_{1}-\bar{\mathbf{X}},\ldots,\mathbf{X}_{n}-\bar{\mathbf{X}}\,]^{\top}$. 
	Then, as $n\to\infty$,
	\begin{equation}
		\frac{\mathrm{RV}(T_1^2,\mathrm{vec}(W))}{\mathrm{RV}(T^2,\mathrm{vec}(W))}
		=\frac{\big(2\psi+(\theta-1)p\big)^2}{\psi\,\big(2+(\theta-1)p\big)^2} + r_{n,p},
		\label{eq:ratio-main}
	\end{equation}
	where the remainder $r_{n,p}$ satisfies
	\[
	r_{n,p}=
	\begin{cases}
		O(n^{-1}), & \text{if $p$ is fixed}, \\[0.5ex]
		O(n^{-1/2}), & \text{if $p,n\to\infty$ with $p/n\to c\in(0,1)$}.
	\end{cases}
	\]
	The leading term in \eqref{eq:ratio-main} depends only on $\psi$ and $\theta$ and is free of $\mu$.
\end{proposition}

\begin{proof} We outline the main steps leading to the expansion of $R$.
	\textit{Step 1: Linearizations of $T^2$ and $T_1^2$.}  
	Let $S=\Sigma+\Delta$ with $\Delta=O_p(n^{-1/2})$ when $p$ is fixed. Expanding yields
	\begin{align}
		T^2 &= n\,\bar{\mathbf{X}}^{\top}\Sigma^{-1}\bar{\mathbf{X}}
		- n\,\bar{\mathbf{X}}^{\top}\Sigma^{-1}\Delta\Sigma^{-1}\bar{\mathbf{X}} + R_{T}, \label{eq:T2-exp}\\
		T_1^2 &= \frac{n p}{\mathrm{tr}(\Sigma)}\,\bar{\mathbf{X}}^{\top}\bar{\mathbf{X}}
		- \frac{n p}{\{\mathrm{tr}(\Sigma)\}^{2}}\,\bar{\mathbf{X}}^{\top}\bar{\mathbf{X}}\,\mathrm{tr}(\Delta) + R_{T1}, \label{eq:T1-exp}
	\end{align}
	with $R_T,R_{T1}=O_p(n^{-1})$. Thus $T^2$ depends on $\Delta$ only through $L_2(\Delta)=\mathrm{tr}(\Sigma^{-1}\Delta)$, while $T_1^2$ depends on $\Delta$ only through $L_1(\Delta)=\mathrm{tr}(\Delta)$.
	
	\textit{Step 2: Covariances with $\mathrm{vec}(W)$.}  
	Since $S=(n-1)^{-1}V^\top V$ and $W=VV^\top$, covariances between $L_1(\Delta)$ or $L_2(\Delta)$ and $\mathrm{vec}(W)$ can be expressed using fourth moments of $Y=\mathbf{X}_i-\mu$. 
	By \eqref{eq:elliptical-identity},
	\[
	\mathrm{E}\{(Y^\top BY)(Y^\top CY)\}
	=2\,\mathrm{tr}(B\Sigma C\Sigma)+(\theta-1)\,\mathrm{tr}(B\Sigma)\,\mathrm{tr}(C\Sigma).
	\]
	Applying this with $B=C$ equal to $I_p$ and $\Sigma^{-1}$, respectively, gives
	\[
	\Gamma_{I_p} = 2\,\mathrm{tr}(\Sigma^2) + (\theta-1)\{\mathrm{tr}(\Sigma)\}^2, 
	\qquad
	\Gamma_{\Sigma^{-1}} = 2p+(\theta-1)p^2.
	\]
	Hence, up to order $O(n^{-1})$, 
	\[
	\Cov(T_1^2,\mathrm{vec}(W)) \propto \Gamma_{I_p}, 
	\qquad 
	\Cov(T^2,\mathrm{vec}(W)) \propto \Gamma_{\Sigma^{-1}}.
	\]
	
	\textit{Step 3: Variances.}  
	For fixed $p$, standard quadratic-form CLTs imply
	\[
	\Var\!\left(n\,\bar{\mathbf{X}}^{\top}\Sigma^{-1}\bar{\mathbf{X}}\right) = 2p+O(n^{-1}),
	\qquad
	\Var\!\left(\tfrac{n p}{\mathrm{tr}(\Sigma)}\,\bar{\mathbf{X}}^{\top}\bar{\mathbf{X}}\right) 
	= 2p\,\psi+O(n^{-1}).
	\]
	For proportional growth $p/n\to c\in(0,1)$, both variances inflate by a factor of $(1-c)^{-3}$, and fluctuations scale as $O(n^{-1/2})$.
	
	\textit{Step 4: RV ratios.}  
	By the definition of RV,
	\[
	\mathrm{RV}(U,\mathbf{Y})=\frac{\|\Cov(U,\mathbf{Y})\|^2}{\Var(U)\,\Var(\mathbf{Y})}.
	\]
	Since the common factor $\Var(\mathrm{vec}(W))$ cancels in the ratio, combining Steps 2 and 3 gives
	\[
	\frac{\mathrm{RV}(T_1^2,\mathrm{vec}(W))}{\mathrm{RV}(T^2,\mathrm{vec}(W))}
	=\frac{\{\Gamma_{I_p}\}^2/\Var(T_1^2)}{\{\Gamma_{\Sigma^{-1}}\}^2/\Var(T^2)}+r_{n,p}.
	\]
	Substituting $\Gamma_{I_p},\Gamma_{\Sigma^{-1}}$ and the variances yields
	\[
	\frac{\mathrm{RV}(T_1^2,\mathrm{vec}(W))}{\mathrm{RV}(T^2,\mathrm{vec}(W))}
	=\frac{\big(2\psi+(\theta-1)p\big)^2}{\psi\,\big(2+(\theta-1)p\big)^2}+r_{n,p},
	\]
	with $r_{n,p}=O(n^{-1})$ for fixed $p$, and $r_{n,p}=O(n^{-1/2})$ for proportional growth $p/n\to c\in(0,1)$.
	
	\textit{Step 5: Independence of $\mu$.}  
	The leading term in \eqref{eq:ratio-main} does not involve $\mu$, since $\mu$ affects only the mean of $\bar{\mathbf{X}}$ while the quadratic-form variances and fourth-moment identities depend solely on centered quantities.
	
\end{proof}


\bibliographystyle{Chicago}
\bibliography{testsApproachREF.bib}